\documentclass[10pt]{article}
\usepackage[utf8]{inputenc}
\usepackage[T1]{fontenc}
\usepackage{authblk} % for multiple affiliations
\usepackage{orcidlink}
\usepackage{graphicx}
\usepackage{amsmath}

\providecommand{\JournalTitle}[1]{#1}
\providecommand{\bibinfo}[2][]{#2}
\providecommand{\doiprefix}{}
\title{Metamaterials and Fluid Flows}

\author[1]{Francesco Avallone\orcidlink{0000-0002-6214-5200}}
\author[1]{Federico Bosia\orcidlink{0000-0002-2886-4519}}
\author[2]{Yi Chen\orcidlink{0000-0002-6614-976X}}
\author[3]{Giada Colombo\orcidlink{0000-0002-6074-3282}}
\author[4]{Richard Craster\orcidlink{0000-0001-9799-9639}}
\author[5]{Jacopo Maria De Ponti\orcidlink{0000-0002-6155-2031}}
\author[6]{Nicol\`o Fabbiane\orcidlink{0000-0003-0010-489X}}
\author[7]{Michael R.~Haberman\orcidlink{0000-0002-7159-9773}}
\author[8]{Mahmoud I. Hussein\orcidlink{0000-0002-6583-1425}}
\author[9]{Wontae Hwang\orcidlink{}}
\author[3]{Umberto Iemma\orcidlink{0000-0002-8940-1489}}
\author[10]{Abigail Juhl\orcidlink{0000-0002-8946-7357}}
\author[11]{Muamer Kadic\orcidlink{0000-0002-4692-5696}}
\author[12]{Marios Kotsonis\orcidlink{0000-0003-0263-3648}}
\author[11]{Vincent Laude\orcidlink{0000-0001-8930-8797}}
\author[13]{Olivier Marquet\orcidlink{0000-0001-7284-6361}}
\author[14]{Fabien Mery\orcidlink{0000-0003-1408-2003}}
\author[12]{Theodoros Michelis\orcidlink{0000-0003-4836-6346}}
\author[15]{Mostafa Nouh\orcidlink{0000-0002-2135-5391}}
\author[12]{Daniele Ragni\orcidlink{}}
\author[16]{Marie Touboul\orcidlink{0000-0003-4052-0994}}
\author[2]{Martin Wegener\orcidlink{0000-0002-9770-2441}}
\author[17]{Anastasiia O. Krushynska\orcidlink{0000-0003-3259-2592}}

\affil[1]{Politecnico di Torino, Department of Applied Science and Technology, Torino, 10129, Italy}
\affil[2]{Institute of Applied Physics and Institute of Nanotechnology, Karlsruhe Institute of Technology (KIT), 76128 Karlsruhe, Germany}
\affil[3]{Roma Tre University, Department of Engineering, Rome, 00146, Italy}
\affil[4]{UMI 2004 Abraham de Moivre-CNRS, Imperial College London, SW7~2AZ, London, UK}
\affil[5]{Politecnico di Milano, Department of Civil and Environmental Engineering, Milan, 20133, Italy}
\affil[6]{DAAA, ONERA, Institut Polytechnique de Paris, Ch\^atillon, 92320, France}
\affil[7]{The University of Texas at Austin, Walker Department of Mechanical Engineering, Austin, TX, 78712-1591, USA}
\affil[8]{University of Colorado Boulder, Smead Department of Aerospace Engineering Sciences, Boulder, Colorado 80303, USA}
\affil[9]{Seoul National University, Department of Mechanical Engineering, Seoul, 08826, Republic of Korea}
\affil[10]{Air Force Research Laboratory, Wright-Patterson AFB, Ohio 45433, USA}
\affil[11]{Université Marie et Louis Pasteur, CNRS, Institut FEMTO-ST, Besançon, 25000, France}
\affil[12]{Delft University of Technology, Faculty of Aerospace Engineering, Delft, 2629HS, The Netherlands}
\affil[13]{DAAA, ONERA, Institut Polytechnique de Paris, Meudon, 92190, France}
\affil[14]{DMPE, ONERA, Université de Toulouse, 31000, Toulouse, France}
\affil[15]{University at Buffalo (SUNY), Department of Mechanical and Aerospace Engineering, Buffalo, NY 14032, USA}
\affil[16]{POEMS, ENSTA, CNRS, INRIA, Institut Polytechnique de Paris, 91120, Palaiseau, France}
\affil[17]{University of Groningen, Faculty of Science and Engineering, Groningen, 9747AG, The Netherlands}
\begin{document}
\maketitle
\begin{abstract}
Understanding and controlling the dynamic interactions between fluid flows and solid materials and structures—a field known as fluid–structure interaction —is central not only to established disciplines such as aerospace and naval engineering but also to emerging technologies such as energy harvesting, soft robotics, and biomedical devices. In recent years, the advent of metamaterials—rationally designed composites with properties beyond their constituents, often not found in conventional materials—has provided exciting opportunities for rethinking and redesigning fluid–structure interaction. The premise of engineering the internal structure of materials interfacing with fluid flows is opening a new horizon for precise and effective manipulation and control of coupled fluidic, acoustic, and elastodynamics responses. This review focuses on this relatively unexplored interdisciplinary theme with broad real-world technological significance. Key performance metrics, such as fuel consumption of transport systems, efficiency of renewable energy extraction, mitigation of noise emissions, and resilience to structural fatigue, depend on the control of interactions between flow, acoustic, and vibration mechanisms. Flow control, for example, which spans a wealth of regimes such as laminar, transitional, turbulent, and unsteady separated flows, is highly influenced by the ability to tailor fluid–structure interaction behavior. We survey and discuss theoretical frameworks that describe the interplay between fluids and elastic solids, with a focus on contemporary work and emerging concepts. The paper is organised into three main sections—flow-structure interactions, acoustic–structure interactions, and exotic metamaterial concepts with potential impact on fluid-structure interaction—and concludes with perspectives on current challenges and future directions in this rapidly expanding area of research.
\end{abstract}

%\flushbottom

\tableofcontents
\clearpage
\section{Introduction}
Dynamic interactions between fluid flows and solid structures underpin various natural phenomena and engineered systems. Known as fluid–structure interaction, they are central not only to traditional fields like aerospace and naval engineering but also to emerging technologies in energy harvesting, soft robotics, and biomedical devices. In recent years, the advent of metamaterials—composites designed to exhibit effective properties that go beyond those of their ingredients, and that are often unique and non-intuitive—has provided exciting opportunities for rethinking and redesigning such interactions. Architecting the topology of materials enables the opportunity to manipulate mechanical, acoustic, and/or fluidic responses, and offer new strategies for controlling the coupled fields.

\begin{figure}[h!]
    \centering
    \includegraphics[width=\linewidth]{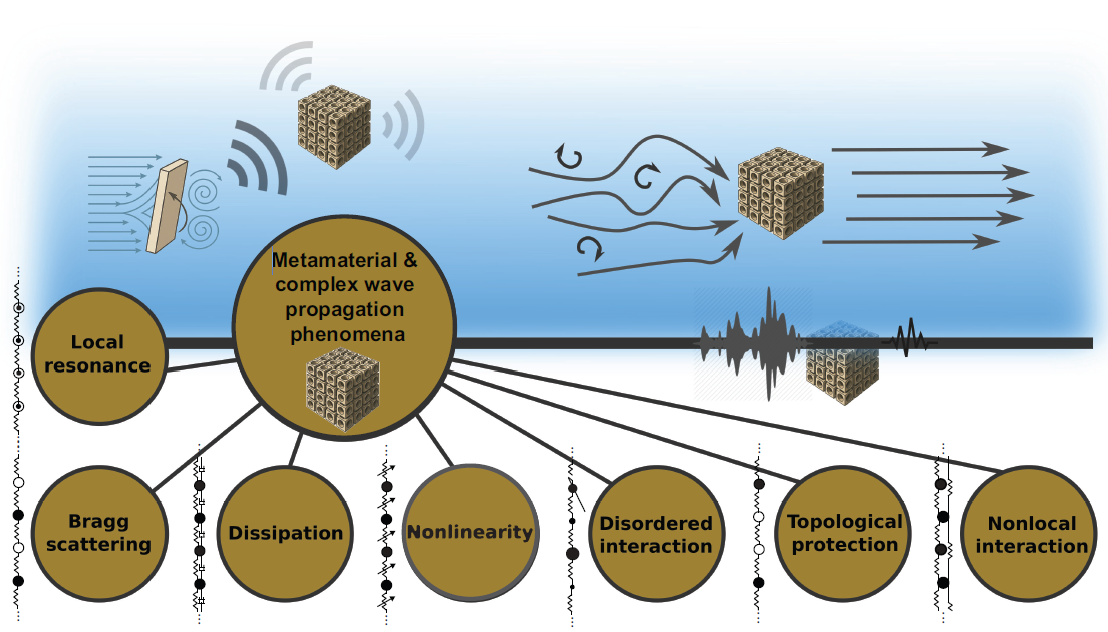}
    \caption{{\bf Metamaterial and complex wave propagation phenomena for fluid-structure interactions.} The top part illustrates coupled flow- and sound-structure coupling with bulk metamaterials, the central part shows an example of flow- or sound-surface interactions, and the bottom part overviews the wave manipulation mechanisms for engineered fluid-structure interactions.}
    \label{fig:1}
\end{figure}

Here, we summarize an under-explored but critically important frontier: the interaction between metamaterials and various wave motion mechanisms that can be engineered at the material level, and fluid (i.e., gas and liquid) flows—as conceptualised in Figure \ref{fig:1}. This focus addresses pressing engineering challenges where controlling or exploiting fluid-structure interaction can dramatically improve performance. Consider, for instance, the hydrodynamic drag experienced by maritime vessels during long-distance travel. A significant portion of the vessel's fuel consumption is dictated by viscous shear forces at the hull–water interface. Similarly, in aeronautical applications, turbulent and/or unsteady air flow over a wing or engine nacelle leads to energy losses, acoustic emissions, and structural fatigue. In both cases, tailoring the fluid–solid interface through specially designed surfaces and subsurfaces can offer substantial gains. 
Fluid–structure interactions are also critical in the context of flow-induced vibrations, where unstable flow patterns can excite resonant structural modes, leading to fatigue or failure, or, conversely, can be used for energy harvesting. Moreover, acoustic metamaterials enable precise manipulation of sound waves in elastic or fluidic media, offering new strategies for vibration damping, noise control, and wave redirection. These interactions are inherently multiphysical and multiscale, requiring multiple frameworks including fluid dynamics, elasticity, acoustics, and interface phenomena.

To provide a cohesive approach to this interdisciplinary domain, we structure the present perspectives paper around three central themes that illustrate the potential of metamaterials and metasurfaces in acousto-fluid-structure interacting systems: 
\begin{itemize}
    \item \textbf {Flow-structure interactions.} We examine how the geometry and material properties of structured materials or patterned surfaces and subsurfaces can control the behaviour of transitional boundary layers, the suppression of turbulence, and the dissipation of energy in internal or external flows. We also touch upon the prospects of metamaterials in manipulating surface gravity waves in liquid flows.
    \item \textbf{Acoustic interactions with structures.} We discuss how metamaterials can manipulate sound generation and propagation, using mechanisms such as local resonance, Bragg scattering, and anisotropic impedance. These concepts are vital for aeroacoustic design and noise mitigation, especially in regimes of high-speed flows. We separately consider ventilated designs attenuating noise in flow regimes and discuss the potential of metamaterials in the manipulation of micro- and nano-particles by sound waves. 
    \item  \textbf{Exotic metamaterial concepts in fluid-structure interaction.} We discuss how striking functionalities of metamaterials, including topologically protected states, nonlocal elasticity, and time- or spatiotemporally-varying properties, can be used to enhance wave control in flow environments. We provide an outlook on how these intriguing metamaterials can open new possibilities and degrees of freedom in flow control by going beyond rigid-wall assumptions and the limits of Cauchy elasticity.    
\end{itemize}

To ground our discussion, we refer to Figure \ref{fig:1}, which presents concepts of fluid–structure interactions. The schematics of the flow– and sound-structure interactions have direct counterparts in real-world contexts. For instance, aircraft wings interact dynamically with surrounding airflow, which influences lift, drag, and structural response, providing a classic example of coupled flow–structure interactions. In addition, the inlets of the aircraft engine are sites of complex flow patterns that couple with rotating and stationary components. These interactions are crucial in determining aircraft performance, fuel efficiency, and resilience to structural fatigue. Another example focuses on more localized flow-structure interactions, such as deformations of a flexible structure in response to fluid flow, a key mechanism in applications, including flow sensors, energy harvesters, and adaptive aerodynamic surfaces. 

These examples illustrate the necessity of advanced theoretical and computational tools capable of describing and predicting acousto-flow-structure interactions. Therefore, in the sections that follow, we develop theoretical frameworks rooted in continuum mechanics, fluid dynamics, and materials science to model these rich and complex interfaces. We survey recent progress in fluid flow control using engineered material and surface/subsurface topologies, and overview the status of metamaterial research in the context of elastic, acoustic, and fluidic environments.

\subsection{Theoretical Framework for Fluid–Structure Interactions in Metamaterials}
\begin{figure}[h!]
    \centering
    \includegraphics[width=\linewidth]{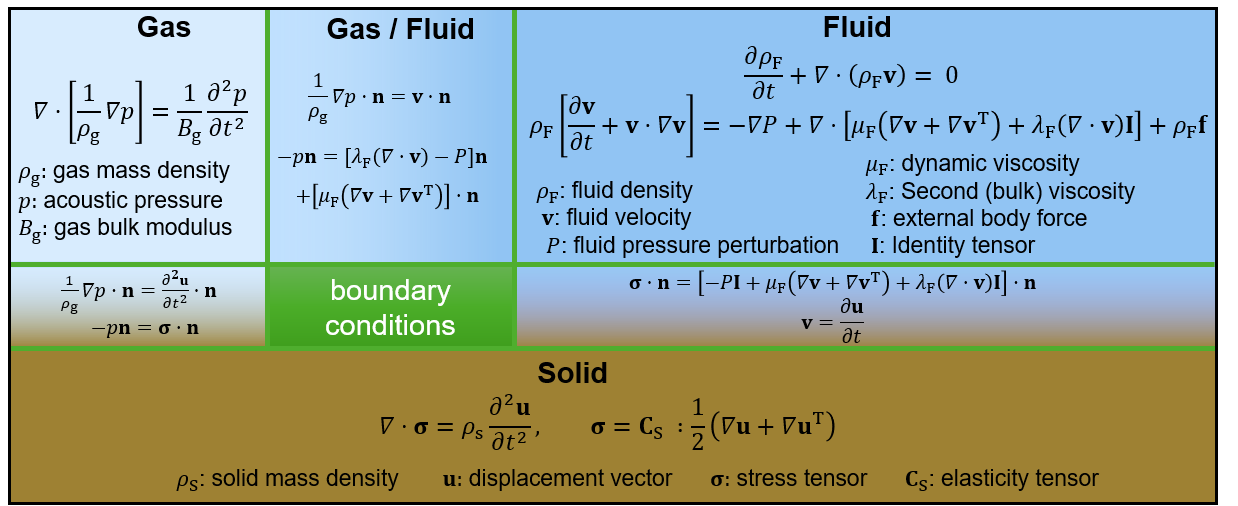}
    \caption{{\bf Illustrative representation of the theoretical formulation for fluid-structure interaction}. The three main dynamic equations for gas, fluid, and solid media, together with the interface boundary conditions for the three different combinations.}
    \label{fig:eq}
\end{figure}

Fluid–structure interaction lies at the core of many physical phenomena, from flow-induced drag on ship hulls to acoustic wave propagation in architected solids. Metamaterials enable novel control over such interactions but necessitate a deep understanding of each physical problem. We now present the governing equations for fluids, solids, and their coupling, forming the theoretical foundation for the remainder of this study. Figure \ref{fig:eq} summarizes the governing equations for gases, fluids, and elastic solids, along with the coupling conditions at their interfaces. Namely, we highlight the linear acoustic wave equation in gases, the Navier-Stokes and Cauchy momentum equations in fluids, and the conservation of linear momentum and linear wave equation in anisotropic solids. We note that the Navier-Stokes and Cauchy momentum equations are applicable in gases when transport and nonlinearities are non-negligible, as is the case in aeroacoustics. More complex elastic wave equations may be necessary to capture motion in elastic metamaterials with structured microstructure that has small-scale asymmetries, such as chirality, to tailor the fluid-solid interaction\cite{maugin2010mechanics,milton2007modifications,frenzel2017three}.
This set of equations provides a unified framework for analyzing how structured media interact with fluids and acoustic waves. The following sections will apply this framework to specific problems. 

In media with periodic structure, such as phononic crystals or acoustic metamaterials, wave propagation is often analyzed using \textit{Bloch's theorem}. This theorem states that the solution to a wave equation in a periodic medium can be written as a Bloch wave—a plane wave modulated by a periodic function. As a result, the analysis can be confined to a single unit cell with Bloch-periodic boundary conditions. Solving the corresponding eigenvalue problem yields the \textit{dispersion relation}, which provides essential information such as the phase and group velocities, as well as potential stop bands or complete band gaps. Fluid flows \cite{tritton2012physical} are typically characterized by the \textit{Reynolds number}, defined as
\[
\text{Re} = \frac{\rho_{\rm F} v L}{\mu_{\rm F}},
\]
where $\rho_{\rm F}$ is the fluid density, $v$ is a characteristic velocity, $L$ is a characteristic length scale, and $\mu_{\rm F}$ is the dynamic viscosity. This dimensionless quantity helps distinguishing between different flow regimes: laminar flow dominates at low Reynolds numbers (viscous forces prevail), while turbulent flow emerges at high Reynolds numbers (inertial forces dominate), with a transient flow regime in between. The knowledge of the flow regime is critical to selecting the appropriate models and analysis methods.

The \textit{finite-element method} is widely used to solve partial differential equations in domains with complex geometries or heterogeneous materials \cite{lomax2001fundamentals}. This method is particularly effective because of its flexibility in handling unstructured meshes, curved boundaries, and spatially varying properties. It also supports multiphysics coupling, making it well-suited for applications such as fluid-structure interaction and acoustic-fluid interface problems. The finite-element method enables local refinement and provides accurate numerical approximations even in challenging domains where analytical solutions are intractable.

In summary, the equations presented in Figure \ref{fig:eq} lead to three main sections in this review paper, organized by application domain, and followed by a final section Conclusion and Perspectives.

\section{Flow-structure interactions}

\begin{figure}[h!]
    \centering
    \includegraphics[width=1\linewidth]{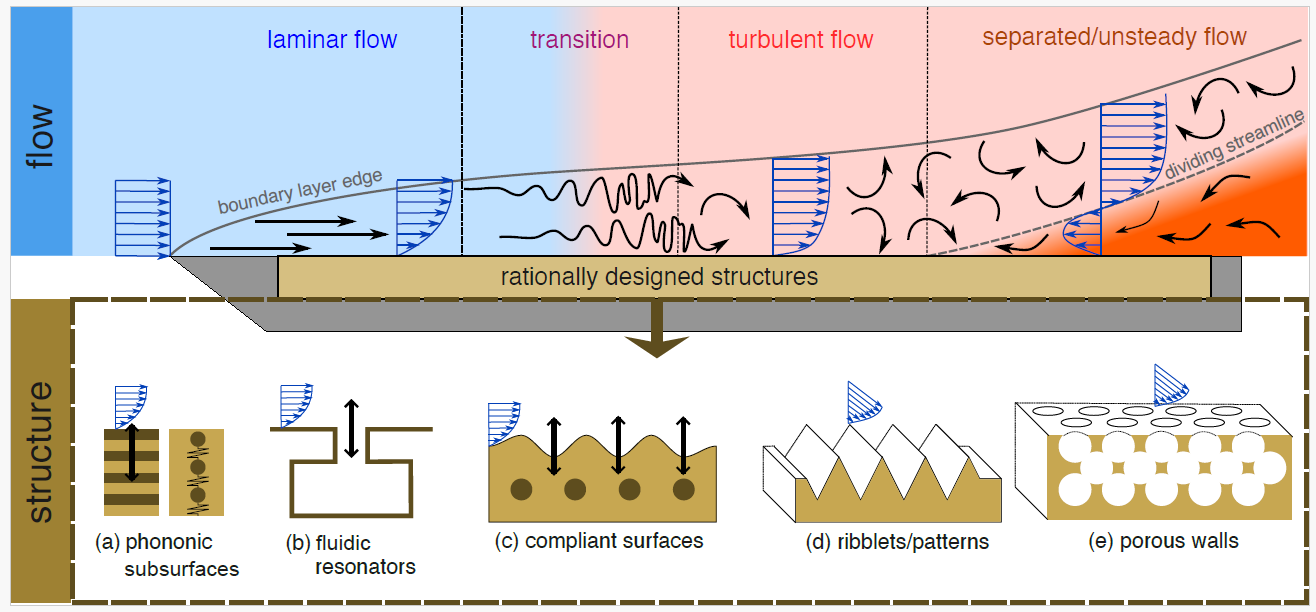}
    \caption{{\bf Conceptualization of typical flow regimes interacting with engineered surfaces/subsurfaces.} (Top) Schematic representation of typical flow regimes in order of increasing unsteadiness. (Bottom) Structural solutions for flow control, (a)  Phononic Subsurfaces (PSubs)~\cite{hussein2015flow,kianfar2023phononic,michelis2023attenuation,willey2023multi,schmidt2025perturbation,navarro2025stabilization}, (b) fluidic resonators~\cite{Michelis2023b,dacome2024small,hassanein2024interaction}, (c) compliant surfaces~\cite{benjamin1960effects,carpenter-et-al:ja2001, Fabbiane2025,pfister:phd2019}, (d) riblets~\cite{walsh1978drag,pollard1998passive,ghaemi2020passive} or patterns~\cite{sharma2020turbulent, Zoppini2022,zoppini2023receptivity,piest2024mitigation,hickey2020targeted,luo2003design,wen2023dynamic},  and (e) porous walls~\cite{fedorov2001stabilization,rosti2018turbulent, Klausmann}.}
    \label{fig:FSI}
\end{figure}

The ability to purposefully engineer the internal structure of materials interfacing with fluids in a manner that allows for target desirable interactions opens up a breadth of opportunities for controlling fluid flows towards improving the responses and overall performance of aero/hydrodynamic systems. Such improvements can manifest as reduction of drag, increase of lift and/or attenuation of unsteady loads otherwise leading to fatigue. The plurality of flow regimes and available concepts of architected structures provides us with a wealth of flow-structure interactions. The salient details of such fluid-structure interactions are specific to the particular combination of architected structure and flow regime, as conceptualised in Figure \ref{fig:FSI}. 

\subsection{Interactions with transitional flow instabilities}
\label{Sec_2_1}

In many engineering flows, the transition from the laminar to the turbulent state is primarily driven by the inception, development, and growth of convective flow instabilities, also known as hydrodynamic instabilities.~\cite{kachanov:arfm1994} A prominent example in two-dimensional (i.e., non-swept wing) flows are Tollmien-Schlichting (TS) waves. The initially linear behaviour of TS wave modes has inspired active control strategies based on the principle of localised perturbation superposition, namely the conscious creation of a "counter response" of the same amplitude and opposite phase as the original TS wave.~\cite{milling1981tollmien} This can be achieved actively via oscillating surface actuation, creating a response normal to the vertical perturbation component of a propagating wave. In the realm of passive flow control, the concept of phononic subsurfaces (PSubs) emerged over the last decade to create the target phased interference in a responsive and functionally effective manner.~\cite{hussein2015flow} A PSub relies on the excitation and management of elastic waves within a phononic material~\cite{hussein2014dynamics} buried beneath the surface, nominally forming a single-input, single-output system. These systems are typically constructed from periodic materials (phononic crystals)~\cite{hussein2015flow, Barnes_2021, michelis2023attenuation} or locally resonant materials (elastic metamaterials),~\cite{kianfar2023phononic, kianfar2023local, hussein2025scatterless} effectively forming a precisely engineered subsurface phonon environment interacting with the flow instabilities. The wave dispersion properties of a PSub, as well as its frequency response function as a truncated finite phononic structure,~\cite{al2023theory,rosa2023material} can be tailored to generate precise control of the amplitude and phase, thereby stabilising or destabilising the TS waves, depending on the target objective. An alternative to phase control, a PSub may be tuned as a phononic diode to extract and trap undesired perturbations in the flow.~\cite{schmidt2025perturbation} A present challenge in exciting PSubs using TS waves is the inherently low amplitude of the latter (typically, $\mathcal{O}$ ($10^{-6}\times{}U_\infty$) during linear growth; $\mathcal{O}$ ($10^{-2}\times{}U_\infty$) when transition onsets), leading to large impedance mismatch at the fluid-structure interface, particularly if the fluid is in gas state (e.g. air). This may potentially be remedied by tuning the PSub properties to simultaneously exhibit a target compliance over and above the desired phononic properties. The impedance mismatch obstacle may also be handled using fluidic/acoustic single-input, single-output systems such as Helmholtz resonators~\cite{Michelis2023b}. In both PSubs and Helmholtz resonators, attenuation of TS instabilities has been demonstrated when surface displacement is out-of-phase relative to the local TS wave pressure footprint. This leads to local stabilization in the regions of the control surface. To achieve downstream transition delay, a multi-input, multi-output PSub system has been demonstrated,~\cite{willey2023multi} and recently the notion of a lattice of PSubs was shown to enable downstream stabilization as well.~\cite{hussein2025scatterless}

Looking back from a historical perspective, early fundamental research on the passive mitigation of TS waves via flexible walls was established in the 1960s~\cite{benjamin:jfm1960,landahl:jfm1962} and comprehensively reviewed in the early 2000s.~\cite{carpenter-et-al:ja2001}
The interaction between the near-wall flow and a compliant wall gives rise to a complex interplay of instabilities, which were first systematically documented and categorised.~\cite{benjamin:jfm1963}
Local stability analysis showed that the flexibility of a compliant wall tends to stabilise TS waves, whereas the structural viscous effects introduce a slight destabilising influence.~\cite{carpenter-garrad:jfm1985} In contrast, the fluid-structural interaction between a compliant wall and the boundary layer was found to be destabilising, as the structural compliance can exacerbate coupled fluid-structural instabilities of Rayleigh-Taylor type.\cite{carpenter-garrad:jfm1986} Multilayered compliant wall configurations have been proposed to engender additional control utility on both the elastic compliance and viscous effects.~\cite{yeo:jfm1988} Recently, more advanced global stability analysis methods, such as transient \cite{tsigklifis-lucey:jfm2017} and forced \cite{pfister-et-al:jfm2022} responses, led towards the optimisation of local material properties in the compliant wall structure,~\cite{pfister:phd2019} uncovering streamwise quasi-periodic patterns that are reminiscent of phononic Bragg scattering behaviour, paving the way to a concept that has been targeted in a recent study to simultaneously tackle a TS wave and a flutter instability.~\cite{Fabbiane2025}

The use of structured materials for controlling flow instabilities was recently shown to extend beyond the case of subsonic TS waves. Periodic surface modifications in the form of shallow cavities have been suggested for stabilising hypersonic boundary layers,~\cite{Zhao2019} achieving a near-zero impedance condition at the wall thus effectively suppressing the growth of the dominant viscous instability, namely the second Mack mode.  In swept-wing boundary layers, the dominant mechanism of transition are crossflow instabilities, which manifest as stationary co-rotating vortices roughly aligned with the free-stream flow.~\cite{Saric2003} Linear superposition and attenuation of stationary crossflow modes was recently achieved by deploying arrays of periodic roughness elements on the swept-wing surface.~\cite{Zoppini2022,zoppini2023receptivity,ivanov2019delay,ustinov2018cross} Such periodic surface architecture can be designed to target specific amplitude, wavelength, and phase characteristics to generate velocity disturbances that interact destructively, thereby attenuating the crossflow instabilities and delaying transition to turbulence.

\subsection{Interactions with unsteady and separated flows}
Beyond the paradigm of laminar flows exhibiting steady behaviour and turbulent flows characterized by chaotic unsteadiness, regimes of (quasi-)deterministic unsteadiness can manifest under specific conditions.
The emergence of these states is often governed by the interplay between inertial and viscous forces, quantitatively assessed via the Reynolds number. Depending on this ratio, such unsteady regimes may either represent the asymptotic flow behaviour or constitute a transitional pathway culminating in turbulence.
Typical examples are wakes of bluff bodies, mixing layers and jets, and separated boundary-layer flows.
Separated flows often follow geometry-induced or externally imposed increases in pressure (i.e., adverse pressure gradient), whereupon the velocity in the boundary layer experiences deceleration and eventual reversal.
A direct consequence of flow separation is a significant increase in boundary layer thickness and subsequent loss of lift and increase in drag. 
Furthermore, separated shear layers are inherently unsteady due to their inflectional shape, leading key parameters like velocity, pressure, and temperature to become temporally transient.
The ability to postpone or mitigate separation in relevant engineering systems such as aircraft, ground transport vehicles, or turbomachinery can significantly reduce drag, enhance lift, and minimise unsteady structural loads \cite{gad1991separation}.

Flow control methods to prevent separation typically involve increasing turbulent mixing at the wall near the separation point by passive or active means (e.g., porous coatings or structures, jets, slats or flaps, plasma actuators, acoustic actuators, morphing architectures).~\cite{ashill2005review,Cattafesta}
Separation can be passively prevented by using distributed vortex generators to induce mixing between upper and lower stratifications within the boundary layer,~\cite{lin2002review} by contouring a structure to decrease the adverse pressure gradients that lead to separation, or by adding patterned geometries (e.g., vanes or riblets) to the surface to increase mixing at the separation point. Plasma actuators based on the dielectric barrier discharge mechanism have been used in various external flow applications, ranging from low speeds to supersonic speeds.~\cite{Corke2010} Synthetic jets have been used to control flow separation by altering the local pressure and vorticity distributions.~\cite{Glezer2002} Porous coatings have been applied to alter the wake structure and reduce drag on a bluff body,~\cite{Klausmann} and recently there have been efforts on decoupling the effects of porosity and permeability for control over the flow structure and pressure drop around porous structures.~\cite{Seol2023, Seol2024}

PSubs, discussed in Section \ref{Sec_2_1}, are also applicable for separation control, whereby the intervention occurs passively by in-phase interferences between the vibratory motion of an elastic surface and the flow perturbations, to cause destabilization.~\cite{hussein2015flow,kianfar2023phononic,kianfar2023local} Some notable efforts for passive separation control include the introduction of microcavities near the leading edge under a wing. These were found to inhibit the bursting of the laminar separation bubble, thereby delaying the onset of dynamic stall.~\cite{visbal2023passive}  Kirigami sheets have been proposed, designed to create an array of tilted surface elements, which could suppress a separation bubble.~\cite{wen2023dynamic} 
Inherent unsteadiness in flow, with or without separation, can be unavoidable in certain scenarios.
In such cases, the primary objective shifts towards mitigating oscillation amplitudes within the flow, thereby attenuating periodic loads imposed on the primary structure.
Recent investigations have explored the potential of architected materials as effective control devices for such applications.
In particular, they have been used to attenuate downstream unsteadiness induced by backward-facing steps,~\cite{piest2024mitigation} and to stabilize oscillations arising from shockwave interactions with boundary layers.~\cite{navarro2025stabilization}

\subsection{Interactions with turbulent flows}

While the previous two sections addressed long-standing and recent developments in the control of transitional and separated flows, the taming of turbulent boundary layers has received its extensive share of foundational and applied research over the past several decades.~ \cite{gad1996modern, abbas2017drag, corke2018active} Despite extensive efforts to promote laminar flow, the high Reynolds number pertinent in many engineering systems (e.g., aircraft, gas/oil pipes, turbomachinery etc.) entails a strong dominance of turbulent boundary layers which exhibit large momentum diffusivity, consequently increasing skin-friction drag and convective heat transfer over surfaces.~\cite{ghaemi2020passive} Upstream turbulent fluctuations are also a known driver of hypersonic shock waves, leading to the onset of severe thermo-acoustic loads that can be detrimental to vehicle integrity as well as sources of acoustic noise. 
Efforts towards control of turbulent flows have ranged from passive to active approaches, both seeking to destructively interfere with the self-regenerating energy production process resulting from sweeps and ejections within a disorderly turbulence sequence.~\cite{warsop2001current} In a turbulent boundary layer, most of the energy production occurs very close to the wall, and as a result, most control efforts have revolved around surface texture modifications or altering the fluid properties in the vicinity of the wall. For example, wall riblets and grooves represent one of the earliest and most commonly studied techniques, showing at times skin-friction-drag reduction by up to 10\%\cite{ghaemi2020passive}. Other passive efforts have included superhydrophobic surfaces and polymer drag reducers, ~\cite{pollard1998passive} systems of micro-canopies, ~\cite{sharma2020turbulent} and concepts of compliant tensegrity fabrics.~\cite{luo2003design} Similarly, active mechanisms such as streamwise aligned plasma actuators for direct drag reduction through Stokes-layer formation have been proposed.~\cite{hehner2019stokes}

The notion of utilizing structured materials that invoke concepts of permeability, resonance, and, potentially, wave dispersion has recently gained attention due to rapid advancements in heterogeneous material fabrication technology, especially those with tunable and graded properties. In this space, wall surfaces with anisotropic porosity, engineered perforations, microcavities, fluidic resonators, baffles, and streamwise slots with a common plenum have all demonstrated promising results.~\cite{rosti2018turbulent, dacome2024small, hassanein2024interaction}  Wave engineering principles, such as PSubs, have been lately investigated for the purpose of attenuating turbulent-induced unsteadiness in high-speed flows.~\cite{navarro2025stabilization} Other studies have examined the impact of wall-normal blowing and suction on turbulent drag in channel flows, simulating subsurface velocities and setting up the problem for subsurface phononic crystals that respond to fluid forces at the wall.~\cite{lin2024control} Alternatively, since the shear force in a wall-bounded flow is directly related to viscosity, wall heating (liquids) or cooling (gases) provides a different path to tackle the same challenge. Motivated by this aspect, wall-aligned heat strips in compressible turbulent channel flows have recently been shown to achieve a notable drag reduction.~\cite{hickey2020targeted} In this approach, “targeted” control, i.e., the intentional targeting of coherent vortical structures to inhibit turbulence production, necessitates a selective spatial arrangement of streamwise heating strips.

\subsection{Surface gravity waves}

Another problem related to engineered interaction with fluid flows is the propagation of surface gravity waves, e.g., water waves, in the presence of solid obstacles or structured boundaries. For linear (small-amplitude) gravity waves, the fluid particle motion is elliptical (circular in deep water, flattened in shallow water), and the particle velocity decays exponentially with depth. Various effects have been demonstrated in this setting, including wave focusing, ~\cite{Bobinski2018BackscatteringChannel} cloaking, ~\cite{Bobinski2015ExperimentalFocusing} superscattering, ~\cite{Sharma2020OnMaterial} and wave guiding.~\cite{Berraquero2013ExperimentalShifter} 
The main strategies adopted so far to manipulate water waves rely on periodically distributed submerged or surface-emerging pillars or periodical depth variations, i.e., variable bathymetry. For example, band gaps for water waves have been obtained using periodic arrays of immersed resonators such as vertical tubes, an effect that the authors compare to a ``negative gravitational acceleration".~\cite{Hu2011NegativeArrays,Hu2013ExperimentalWaves} Metasurfaces have been effectively used to generate band gaps, for example, using a periodically drilled bottom,~\cite{Hu2003CompleteBottom} or activating Bragg scattering by arrays of floating plates and/or submerged trenches.~\cite{Kar2020BraggTrenches} Numerical models have been adopted to describe the interaction of periodic structures with water waves, predicting refraction,~\cite{Hu2005RefractionArrays} scattering,~ \cite{Kar2020BraggTrenches}, or rainbow effects in spatially graded arrays.~\cite{Bennetts2018GradedWaves, Wilks2022RainbowBarriers} Other experimental studies have demonstrated reflectionless waveguiding in curved channels realized with a layered metamaterial structure.~\cite{Archer2020ExperimentalArrays} On the other hand, relatively few studies have considered the possibility of employing periodic arrangements of submerged resonators to control water waves, partially due to the complexity of the problem involving oscillations of the resonators coupled with their movement in a fluid, with correlated surface and viscous effects. Recently, the concept of an inverted subsurface pendula has been introduced to couple both local resonance and Bragg effects and generate band gaps in water waves.~\cite{DeVita2021AttenuatingMetamaterials, DeVita2021AWaves}

\section{Acoustic interactions with structures}
In many practical applications (e.g., in aerospace, energy, and turbo-machinery systems) involving acoustic metamaterials, the presence of background flows cannot be neglected. The interaction of acoustic waves with both the surrounding structures and flow determines how sound is generated, transmitted, or suppressed. This section focuses on three main categories relevant to acoustics and flow: noise generation and mitigation in aerodynamics (Section~3.1), ventilated metamaterials for flow-noise mitigation (Section~3.2), and particle manipulation in fluids (Section~3.3).

\begin{figure}[h!]
    \centering
    \includegraphics[width=\linewidth]{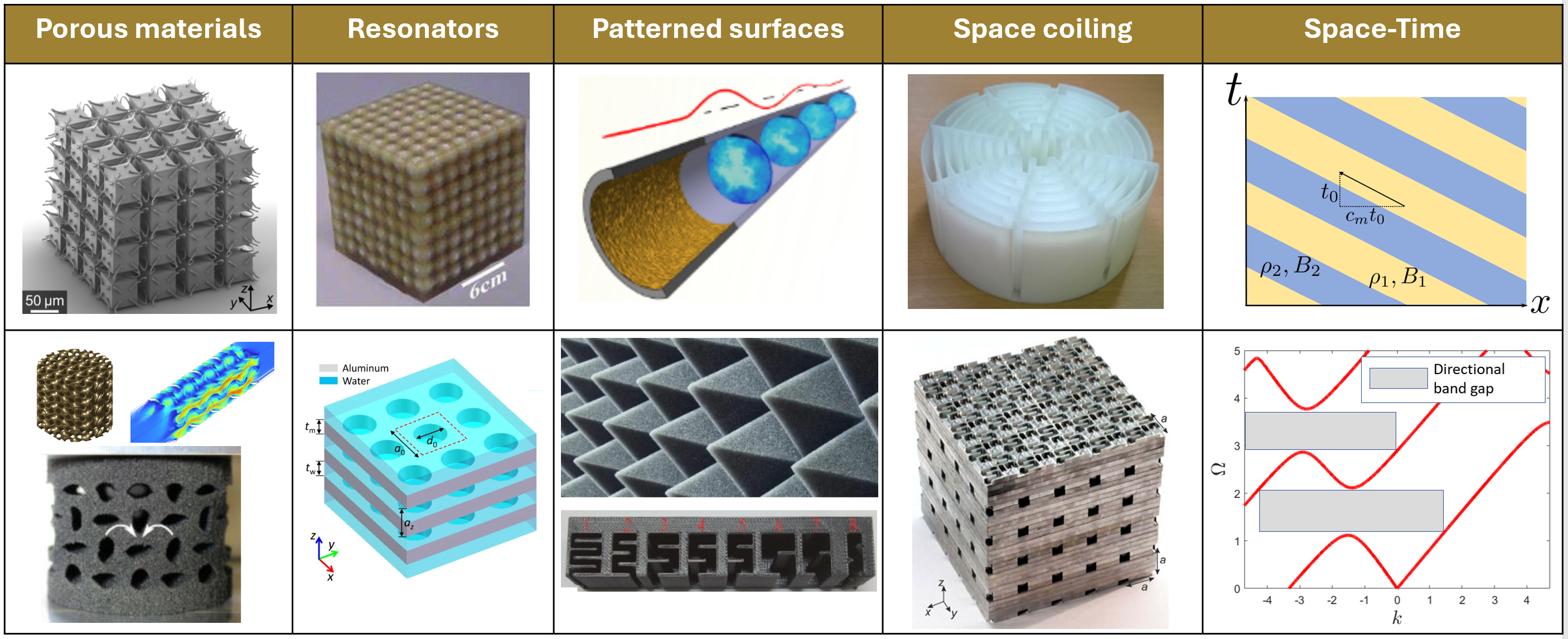}
    \caption{ {\bf Examples of metamaterials for controlling acoustic waves and flow regimes.}
    {\bf Poroelastic metamaterials} exhibit effective negative compressibility when subjected to an increase in external air pressure.~\cite{qu2018three} Porous inserts can be used to reduce internal flow turbulence for decades, while triply-periodic minimal structures deliver more predictable turbulence reduction.~\cite{piest2024mitigation}
    {\bf Local resonators:} acoustic metamaterials with local resonators operate in subwavelength regimes by controlling sound waves of wavelengths much larger than their dimensions. (top image is from Ref.~\cite{liu2000locally}) 
    {\bf Patterned surfaces}, including riblets, viscous coatings (top), and periodic inserts, are widely applied to mitigate turbulence in pipes,~\cite{van2020turbulent}, while corrugated walls and labyrinthine metasurfaces (bottom) are excellent absorbers of broadband low-frequency sound.~\cite{liang2018wavefront}
    {\bf Space coiling:} labyrinthine metamaterials characterized by complex internal geometries that manipulate wave
propagation through extreme path lengthening.~\cite{cheng2015ultra, frenzel2013three}
    {\bf Spatiotemporally-periodic bi-materials}. (top): Space-time property map with the properties (e.g., the mass density and the bulk modulus) with wave-like periodic dependence on both space and time and modulation speed  $c_m$. (bottom): Asymmetric dispersion diagram presenting directional band gaps in the case of subsonic modulation ($0<c_m<\mathrm{min}(c_1,c_2)$).}
    \label{fig:3}
\end{figure}

\subsection{Aerodynamic generated noise}
In aeronautics, acoustic metamaterials have been proposed as noise-control technologies to reduce surface pressure fluctuations and suppress aeroacoustic sources.~\cite{Palma2018} Applications include trailing-edge noise mitigation in wind turbines,~\cite{Geyer2010, RubioCarpio2019, Teruna2020} noise due to turbulence–surface interaction in engines and propellers,~\cite{teruna2021, Jawahar2023,Avallone2018} and landing-gear noise reduction.~\cite{Zamponi2025} Flow-permeable surfaces such as porous liners or engineered perforations are common solutions,~\cite{Ross2024, Tek1957}. However, their acoustic response depends on parameters like permeability and form coefficients, which vary significantly under grazing flow where convection dominates.~\cite{Luesutthiviboon2021, Johnson1987, iemma2020} These interactions between flow and metamaterial structures (which are usually rough surfaces) generate vorticity and alter turbulence, thereby complicating both prediction and design of improved acoustic liners.~\cite{zamponi2024}
To investigate these effects, many research units have developed test rigs to study acoustic liners, including metamaterial concepts, under grazing flow conditions.~\cite{jones2005benchmark,bonomo,auregan2008,zheng2022, minotti} Although differing in flow generation and acoustic excitation characteristics, sample sizes, and working frequencies, these facilities are designed to determine the acoustic impedance representing the metamaterial response using either direct or indirect methods. Facilities that employ direct methods measure the axial wave number, thereby allowing the determination of impedance without iteration.~\cite{jing2008} Indirect approaches determine the impedance using non-intrusive wall pressure measurements combined with inverse modelling of the convected Helmholtz equations and comparison with measurements.~\cite{jones2005benchmark, eversman2011,primus2013adjoint, Ingard1959,Myers1980OnFlow} Refinements include accounting for shear flow with Pridmore-Brown formulations~\cite{jones2010effects} or employing advanced diagnostics such as laser Doppler velocimetry to measure the acoustic velocity field.~\cite{primus2013adjoint}
Only a few studies have examined the issue of uncertainty quantification in the impedance education process with grazing flow.~\cite{Quintino2025} Monte Carlo~\cite{brown} and Bayesian~\cite{rroncen2018} approaches have been applied to assess sensitivity to flow conditions (e.g., Mach number ).~\cite{zhou} The drag penalty associated with acoustic liners has also become a central concern, which must be a key consideration for metamaterial solutions. Experiments have confirmed that coatings with perforated resistive layers can increase viscous drag when subjected to strong acoustic excitation, thus emphasizing the need for a trade-off between acoustic benefit and aerodynamic penalty.~\cite{howerton2017,jasinski} Complementary studies using infrared thermography and micro-particle image velocimetry have provided further insight into local flow–liner interactions and their thermal and mechanical coupling.~\cite{leon2019,lafont2021}

\subsection{Ventilated metamaterials}
A promising use of acoustic metamaterials is to advance noise control technology for applications that require flow. This family of metamaterials is commonly referred to as ventilated acoustic metamaterials.~\cite{zhen2025resonance,kumar2019present} Classical examples of this category of applications include noise control in heating, ventilation, and air conditioning systems for residential and office spaces, sound generation of high-speed flows in piping of industrial equipment, and turbomachinery noise, including jet engine noise.~\cite{ingard2009noise} Traditional solutions to these problems, such as branch Helmholtz or quarter-wavelength resonators, expansion and plenum chambers, mufflers~\cite{ingard2009noise}, and Herschel--Quincke tubes~\cite{stewart1926tube} and similar branching structures~\cite{poirier2011use,papathanasiou2022herschel} provide narrow-band noise isolation with minimal disruption to fluid flow. These architected acoustic wave filters predate metamaterial solutions by many decades and may, in fact, among others, be viewed as precursors to recent metamaterial research. Broadband acoustic absorbers in current industrial applications consist primarily of acoustic liners constructed from porous media or micro-perforated structures, both of which provide reduced acoustic isolation when compared to resonant structures and can restrict flow.~\cite{ingard2009noise,ko1971sound,arenas2010recent} 

Ventilated acoustic metamaterials represent a merging of new scientific concepts and fabrication capabilities with classical solutions to acoustical problems to address longstanding noise control problems.~\cite{zhen2025resonance,kumar2019present} These metamaterial solutions are generally constructed of two components: (i) one region that allows the flow of gas or fluid and (ii) another that contains subwavelength geometries to generate high acoustic isolation via resonance.~\cite{zhen2025resonance} These metamaterial structures can be made geometrically compact by leveraging coiled-space designs~\cite{liang2012extreme} of various form factors, including labyrinthine~\cite{ghaffarivardavagh2019ultra,sun2020broadband,tang2022broadband,zhu2023nonlocal} and helical~\cite{liu2017spiral,xiao2021ventilated,chen2022broadband,lee2025compact} architectures arranged at duct boundaries or at interfaces between domains through which fluid flow is desired. This approach has been generalized to create structures that surround acoustic sources in free space to provide acoustic isolation while allowing flow through the volume.~\cite{liu2021three,krasikova2023metahouse} The key attribute of ventilated acoustic metamaterials is the ability to design performance in an acoustically compact space, enabling tailored acoustic isolation for target broadband~\cite{meng2023subwavelength,yang2025phase,xiao2021ventilated,tang2022broadband,yang2017optimal} and even tunable~\cite{li2022frequency,wen2023origami} isolation. Ventilated acoustic metamaterials therefore represent a highly relevant and practical solution for applications where flow is an essential feature of system functionality and where noise control is required.

\subsection{Particle manipulation by acoustic field}
Controlled manipulation of particles at the microscale and nanoscale is essential across a wide range of scientific and technological fields, from particle physics to biomedicine. It plays a critical role in studying biological processes at the single-cell or subcellular vesicle level, identifying protein biomarkers associated with specific diseases, assembling DNA, and understanding fundamental particle dynamics.~\cite{zhang-et-al:jnm2020} Microfluidic technologies have emerged as powerful tools for particle manipulation, enabling transportation, separation, trapping, and enrichment through various mechanisms, including hydrodynamic, acoustic, electrical, optical, and magnetic methods. However, due to the extremely small size of nanoparticles, direct mechanical manipulation is often impractical. As a result, non-contact methods—typically involving dynamic control of external gradient force fields—are preferred. 
Among different methods, acoustofluidics~\cite{godary-et-al:acs2025}—the use of sound waves to exert forces within fluidic environments—has gained significant attention. This approach is particularly attractive due to its non-contact, label-free nature, high biocompatibility, and the ability to manipulate particles over relatively large volumes. These systems commonly employ surface acoustic waves~\cite{ding-et-al:loc2016, guo-et-al:pnas2016} or bulk acoustic waves,~\cite{wu-et-al:micronano2019} which are typically employed to achieve precise control over particles in acoustic tweezers.~\cite{shen-et-al:sciadv2024} Acoustofluidic devices manipulate fluids and particles using acoustic radiation forces and streaming. When a traveling surface acoustic wave encounters a liquid, its higher viscosity causes part of the wave to refract as a longitudinal wave, forming a “leaky” or “pseudo” surface or bulk acoustic wave. Reflections at the liquid boundaries induce acoustic streaming—a nonlinear effect that converts wave attenuation into steady fluid flow, enabling manipulation of suspended particles.~\cite{ding-et-al:loc2016}
Although acoustic radiation forces enable precise control, their strength is inversely proportional to both the fluidic channel dimensions and the acoustic wavelength.~\cite{bruus:loac2013} Consequently, achieving high-resolution manipulation at the micrometer scale typically requires high-frequency interdigital transducers and complex electronic systems, which are costly and technically demanding. Recent advances in additive manufacturing offer novel solutions to these limitations,~\cite{ellen:loac2018} enabling the fabrication of corrugated surfaces and phononic crystals for advanced particle manipulation.~\cite{fei:prappl2020, huang:apl2022} In particular, acoustofluidic metasurfaces composed of subwavelength inclusions, pillars or other rationally designed composites show strong potential for fluid manipulation and for trapping particles at scales well below the acoustic wavelength~\cite{zeighami:philtransA2024, surappa:natcommun2025}; we envision that combining diverse metasurface geometries will enable the generation of even more complex acoustic fields and fluid flows—without relying on intricate interdigital transducers arrangements.

\section{Exotic elastic metamaterials in fluid-structure interactions}

Elastic metamaterials provide a powerful platform for controlling fluid-structure interaction by coupling structural vibrations with pressure waves in gases and liquids. Their engineered architecture allows for band gaps, local resonances, and anisotropic responses that help suppress turbulence, reduce drag, and dampen flow-induced vibrations.
Classical elastic materials are modelled by Cauchy elasticity—an extension of Hooke’s law to continua—where matter is treated as being composed of infinitesimally small volume elements that can be described by purely translational displacements. This approximation holds when atomic dimensions are negligible compared to vibration wavelengths. In contrast, metamaterials have much larger unit cells, enabling vibrational resonances at relevant frequencies~\cite{kadic20193d} and Bragg scattering with phononic band gaps.~\cite{jin20252024} Unit cell deformations introduce behavior beyond Cauchy elasticity \cite{frenzel2017three}, requiring generalized frameworks such as micropolar elasticity.~\cite{chen2020mapping} Long-range coupling between units can induce nonlocal effects,~\cite{chen2021roton}, and modulating material properties in both space and time enables symmetry-breaking and non-reciprocal behavior.~\cite{craster2023mechanical,chen2025nonlocal,nassar2020nonreciprocity}
Most mechanical metamaterials comprise solid matrices with internal voids—often assumed to be vacuum. In reality, these voids are filled with air or water, allowing weak but significant coupling between solid elastic and fluid pressure waves.~\cite{chen2020high} This interaction, though often neglected due to impedance mismatch, creates opportunities for wave control. In airborne or waterborne media, rigid-wall assumptions~\cite{cummer2016controlling} dominate, but poroelastic metamaterials demonstrate active fluid–structure coupling.~\cite{qu2018three,zhang2021wave}
The following subsections explore how topological, nonlocal, and time-varying metamaterials may be used to enhance wave control in flow environments.

\subsection{Topological Interactions}
Acoustic and elastic topological systems,~\cite{fleury:natcommun2016, huber:natphys2016} like their electromagnetic counterparts,~\cite{khanikaev:natmat2013} exhibit bulk–edge correspondence, where the periodic band structure determines the presence of robust edge or surface modes, insensitive to defects and disorder, and immune to backscattering.~\cite{BertoldiPRL2015, MaNatRevPhys2019}  Underpinning the topological nature of the Bloch bands are associated invariants, which characterize the geometric phase, that is, the phase change associated with a continuous adiabatic deformation of the system; most notably, the Berry phase~\cite{berry:philtrans1984} and its one-dimensional counterpart, the Zak phase.~\cite{zak:prl1989} Many topological insulators have been translated into wave physics using Lieb, kagome lattice \cite{jiang:prb2019} or glide-reflection symmetric crystal interfaces.~ \cite{martínez:prb2022} These points enable the formation of edge states at interfaces between topologically distinct media and have been adapted for waveguiding, localization, and sensing.~\cite{deponti:natcommun2024} More recently, interest has shifted to so-called higher-order topological insulators and semimetals, a novel class of structures where topologically protected modes emerge at lower-dimensional boundaries such as corners or hinges, enabling robust control of sound and mechanical waves beyond conventional topological phases.~\cite{XueNatMat2019, NiNatMAt2019} These materials are said to generalize the conventional bulk-boundary correspondence, recognizing that nontrivial bulk topology can manifest not only at edges but at lower-dimensional features.~\cite{LuoNatMat2021, MaPRL2024} 
Many experimental demonstrations of topological phononic systems have used acoustic waves propagating in the air, mostly neglecting the interaction with the solid skeleton that surrounds the wave system.~\cite{yang2015topological,lu2016valley,lu2017observation,zhang2018topological}
Topological edge states have also been considered in water wave systems that support highly dispersive wave propagation, in the gravity capillary wave regime~\cite{laforge2019observation,makwana2020experimental}
In the case of crystal structures composed of steel inclusions immersed in water, it was found that fluid-structure interaction must be taken into account to faithfully describe the propagation of acoustic waves, and especially the frequency position of Dirac points.~\cite{laforge2021acoustic}
To date, topological wave systems have thus primarily been explored in single-phase media (solid or fluid), with limited understanding of the complex interactions that can occur between a vibrating structure and fluid dynamics. One notable exception is the demonstration of a novel approach to topological phononics by exploiting the interplay between fluid-borne and solid-borne sound waves, leading to the experimental realization of type-II nodal rings in a simple three-dimensional phononic crystal.~ \cite{WuNatComm2022} Through combined theoretical and experimental analysis, the authors reveal that fluid-solid interactions enable strongly tilted drumhead surface states, offering a robust and easily implementable platform to explore advanced topological phenomena in flow-coupled systems. Another recent demonstration of on-chip topological acoustofluidics reveals the complex interplay between phonon valleys, nonlinear fluid dynamics, and the crystallographic structure of the substrate within a microfluidic environment.~\cite{zhao:natmater2025} The study reports the formation of topological pressure nanowells that enable DNA molecule concentration via confined edge modes, offering a new approach to probing biological behavior in fluid and solid media. It also highlights the orientation dependence of acoustofluidic edge states on substrate crystallography. These findings are expected to open new avenues for the integration of topological materials with life sciences and the exploration of topological phenomena in complex media.

\subsection{Local and Nonlocal Interactions}
The Cauchy elasticity equations shown in the green part of Figure 2 are local: stress at one point produces strain only at that point.~\cite{eringen2003nonlocal} This assumption relies on modelling materials by infinitesimally small volume elements, which breaks down in metamaterials with finite-sized unit cells. In contrast, \textit{nonlocal elasticity} allows the strain at one point to depend on stresses elsewhere, modelled through spatial dispersion in reciprocal space or convolution in real space.~\cite{landau2013electrodynamics}
Nonlocal effects in metamaterials arise from higher-order interactions, soft modes, chirality, or time-dependent couplings, as reviewed elsewhere.~\cite{chen2025nonlocal} In the static regime, they are linked to decaying Bloch modes or “frozen phonons”,~\cite{krushynska2011normal,chen2024anomalous} whose long decay lengths produce size effects that can exceed tens of unit cells. Under specific boundary conditions, such structures can exhibit responses that deviate significantly from the assumptions of classical elasticity.
Interestingly, static nonlocal elasticity maps mathematically to Ohm’s law in incompressible electron fluids and laminar fluid flow in networks via Hagen–Poiseuille’s law. This analogy has revealed the emergence of backward currents—flows that oppose the primary direction due to oscillating solutions of the Bloch problem.~\cite{iglesias2025nonlocal} These induce laminar-scale vortices, mimicking effects normally associated with turbulence, and may offer new pathways for fluid manipulation in metamaterial-based channels.
In dynamics, nonlocal metamaterials can produce roton-like dispersion with backward-propagating phonons,~\cite{iglesias2021experimental} enabling precise wave shaping. Through Fourier synthesis, arbitrary dispersion relations can be engineered by tuning beyond-nearest-neighbor interactions.~\cite{wang2022nonlocal,kazemi2023drawing}
Poroelastic metamaterials, which incorporate fluid-filled voids in elastic matrices, are especially relevant here. They enable strong coupling between pressure waves and solid deformation, unlike rigid-wall assumptions.~\cite{qu2018three,zhang2021wave}

\subsection{Space-Time Materials}
Space-time metamaterials, often referred to as spatiotemporally modulated metamaterials, are a class of metamaterials whose properties are modulated not only in space but also dynamically in time, enabling unprecedented control over wave propagation. This combined space-time modulation breaks traditional physical symmetries such as reciprocity and gives rise to a range of exotic wave phenomena. These include non-reciprocal wave propagation, where waves can travel in one direction but are blocked in the opposite one, asymmetric transmission, unidirectional amplification, and frequency and wavenumber conversion. Notably, spatiotemporal modulation generates these phenomena without the need for magnetic-field biasing or material nonlinearity. Such properties have opened new frontiers in designing devices for wave manipulation in acoustics,~\cite{fleury2014nonreciprocity} elastodynamics,~\cite{nassar2020nonreciprocity, Goldsberry2022} electromagnetics,~\cite{galiffi2022photonics} surface Rayleigh waves,~\cite{palermo2020surface} diffusion,~ \cite{Torrent2018,Kang2022} flexural waves,~\cite{Goldsberry2025}  flexural-gravity waves,~\cite{farhat2021spacetime} water waves,~\cite{Koukouraki2024} with potential applications in isolation, sensing, communication, and vibration control. Many works have focused on the case of wave-like space-time modulations of the physical properties~\cite{Cassedy1963,TessierBrothelande2023,Harwood2024} as indicated in the space-time property map shown in Figure~\ref{fig:3} (Space-Time panels). Effective-medium models have been derived for this case, yielding non-reciprocal Willis-coupling, or bi-anisotropic models, if both the density and modulus are modulated in phase.~\cite{Nassar2017,Huidobro2021,Touboul2024} As a result, waves propagating in space-time modulated media break $k \leftrightarrow -k$ symmetry. In parallel, other studies have explored the role of time-modulated boundary conditions,~\cite{Zhu2020} time-modulated interfaces~\cite{Tirole2024} or time-modulated resonators embedded in a quiescent background medium~\cite{Santini2024,Ammari2024} to achieve similar symmetry-breaking effects. Additional uses of space-time modulation at domain boundaries or interfaces between domains include non-reciprocal scattering, pseudo-Doppler effects, and space-time encoding of scattered fields as previously studied in electromagnetic metasurfaces.~\cite{Taravati2022}

At present, very few works combine the concept of space-time modulated metamaterials and fluid flows, although it is notable that there is a well-established community using active control and feedback to delay the transition to turbulence, lower drag and wing loads, and optimise aerodynamic performance.~\cite{Collis2014, Brunton2015} Time-varying metamaterial concepts are beginning to be applied to flow control: for example, using active subsurface diodes to extract undesirable flow perturbations.~\cite{schmidt2025perturbation} The introduction of time variation of material properties or boundary conditions on the same timescale as the disturbance wave period introduces a new degree of freedom and the possibility of wave amplification, decay, or non-reciprocity to additional flow control.

Accurate modelling of these effects is crucial for designing metadevices in dynamic environments. A promising approach reformulates propagation in space-time, treating background flow as local space-time curvature, and exploiting General Relativity tools to better understand acoustic waves in moving media. By linking the aerodynamic flow to a differential operator on a curved manifold, the governing equations gain an invariant form.~\cite{Online:Visser1993} This enables space-time coordinate transformations that embed background flow into metacontinuum designs, yielding solutions driven solely by the convective patterns of the space-time mapping, independent of the targeted acoustic mirage.~\cite{iemma2020} This approach strengthens the theoretical understanding of propagation phenomena and enables the design of advanced aeroacoustic devices based on exotic continua, although incomplete space-time transformations limit effectiveness, making wave control dependent on flow intensity and source position relative to the treated zone.~\cite{Article:GiadaWiley,  Article:GiadaScientificReport}

\section{Conclusion and Perspectives}
This review and perspectives article has discussed the rich and multifaceted field of coupling of material architecture with fluid flows, highlighting a range of distinctive metamaterial principles, and emphasizing recent advances in theory, experiment, and application. We have examined how structured materials—engineered at micro- to macro-scales—can modulate, suppress, or amplify flow-related phenomena such as boundary- or shear-layer instabilities, surface acoustic emissions, and flow-induced structural vibrations. Across laminar, transitional, and turbulent regimes,  phonon engineering of flow-bounded walls are showing great promise for controlling boundary-layer development, delaying separation, and minimizing drag. In acoustics, flow-permeable materials and aeroacoustic liners are being successfully adapted to reduce broadband noise and manage convective effects in complex environments. Emerging metamaterial concepts are enabling localized vibration control and energy trapping, particularly through resonant and topological mechanisms. The incorporation of engineered elastic, porous, and patterned interfaces has begun to shift the paradigm of flow control from empirical surface treatments to physics-informed design of functional surfaces and subsurfaces.

What distinguishes metamaterials and material engineering concepts from conventional materials in this context is the capacity to enable wave-based control strategies—leveraging dispersion, periodicity, local resonance, impedance mismatch, and anisotropy to manage fluid-solid coupling and flow instabilities at a fundamental level. Through approaches like phononic subsurfaces, diodic structures, and compliant phononic-crystal coatings, researchers have demonstrated unique passive and semi-active methods for managing flow behaviour in various systems, including channels, aerofoils, inlets, cavities, and underwater bodies. Although integration into large-scale applications remains challenging, the theoretical foundation and proof-of-concept results offer a compelling path forward.

Looking ahead, integration of cutting-edge metamaterial principles promises to unlock powerful new fluid-structure interaction capabilities. Expanding the use of nonlocal, anisotropic, and topological effects may enable control beyond the current classes of surface interactions, targeting complex disturbance propagation within fluid or structural domains—including in high-Reynolds-number flows. Programmable and reconfigurable metamaterials hold promise for creating flow-responsive systems that adapt in real time to changing conditions, enabling active control of boundary layers, flow separation, and acoustic noise. In addition, multiphase and poroelastic materials with graded impedance or compliance at the fluid-solid interface could provide smoother and efficient control strategies. Incorporating time-varying material properties may break reciprocity limits and harness flow instabilities for energy extraction.

Realizing these opportunities will require tightly integrated modelling, fabrication, and experimentation, and, critically, close collaboration between the fluids and structural dynamics communities. A forward-looking prospect is to move beyond passive structures toward metamaterials that actively shape fluid behaviour—enhancing performance and efficiency in aerospace, marine, energy, biomedical applications, and other systems.

\section*{Acknowledgements }
We acknowledge Euromech659 "Metamaterials in Fluid Flows, Aeroacoustics, and beyond" held in Groningen, The Netherlands, on 25-28 March 2025 as the driving event for this work. M.W. acknowledges support by Deutsche Forschungsgemeinschaft through the Excellence Cluster "3D Matter Made to Order" (EXC 2082). A.O.K. acknowledges financial support by the Dutch Research Council (NWO) through the "MetaFlow" project (KICH1.ST04.22.010) for organizing the Euromech659 event. M.K. acknowledges support by the European Research Council (ERC, MetaWing, 16762850) and Dutch Research Council NWO-I (OCENW.M20.186). F.A. is co-funded by the European Union (ERC, LINING, 101075903). Views and opinions expressed are however those of the author(s) only and do not necessarily reflect those of the European Union or the European Research Council. Neither the European Union nor the granting authority can be held responsible for them.

\section*{Author contributions statement}
All authors wrote and reviewed the manuscript.


\begin{thebibliography}{100}
\urlstyle{rm}
\expandafter\ifx\csname url\endcsname\relax
  \def\url#1{\texttt{#1}}\fi
\expandafter\ifx\csname urlprefix\endcsname\relax\def\urlprefix{URL }\fi
\expandafter\ifx\csname doiprefix\endcsname\relax\def\doiprefix{DOI: }\fi
\providecommand{\bibinfo}[2]{#2}
\providecommand{\eprint}[2][]{\url{#2}}

\bibitem{maugin2010mechanics}
\bibinfo{author}{Maugin, G.~A.} \& \bibinfo{author}{Metrikine, A.~V.}
\newblock \emph{\bibinfo{title}{Mechanics of generalized continua}}.
\newblock Advances in Mechanics and Mathematics (\bibinfo{publisher}{Springer US}, \bibinfo{year}{2010}).

\bibitem{milton2007modifications}
\bibinfo{author}{Milton, G.~W.} \& \bibinfo{author}{Willis, J.~R.}
\newblock \bibinfo{journal}{\bibinfo{title}{On modifications of newton's second law and linear continuum elastodynamics}}.
\newblock {\emph{\JournalTitle{Proceedings of the Royal Society A: Mathematical, Physical and Engineering Sciences}}} \textbf{\bibinfo{volume}{463}}, \bibinfo{pages}{855--880} (\bibinfo{year}{2007}).

\bibitem{frenzel2017three}
\bibinfo{author}{Frenzel, T.}, \bibinfo{author}{Kadic, M.} \& \bibinfo{author}{Wegener, M.}
\newblock \bibinfo{journal}{\bibinfo{title}{Three-dimensional mechanical metamaterials with a twist}}.
\newblock {\emph{\JournalTitle{Science}}} \textbf{\bibinfo{volume}{358}}, \bibinfo{pages}{1072--1074} (\bibinfo{year}{2017}).

\bibitem{tritton2012physical}
\bibinfo{author}{Tritton, D.~J.}
\newblock \emph{\bibinfo{title}{Physical fluid dynamics}} (\bibinfo{publisher}{Springer Science \& Business Media}, \bibinfo{year}{2012}).

\bibitem{lomax2001fundamentals}
\bibinfo{author}{Lomax, H.}, \bibinfo{author}{Pulliam, T.~H.}, \bibinfo{author}{Zingg, D.~W.}, \bibinfo{author}{Pulliam, T.~H.} \& \bibinfo{author}{Zingg, D.~W.}
\newblock \emph{\bibinfo{title}{Fundamentals of computational fluid dynamics}}, vol. \bibinfo{volume}{246} (\bibinfo{publisher}{Springer}, \bibinfo{year}{2001}).

\bibitem{hussein2015flow}
\bibinfo{author}{Hussein, M.~I.}, \bibinfo{author}{Biringen, S.}, \bibinfo{author}{Bilal, O.~R.} \& \bibinfo{author}{Kucala, A.}
\newblock \bibinfo{journal}{\bibinfo{title}{Flow stabilization by subsurface phonons}}.
\newblock {\emph{\JournalTitle{Proceedings of the Royal Society A}}} \textbf{\bibinfo{volume}{471}}, \bibinfo{pages}{20140928} (\bibinfo{year}{2015}).

\bibitem{kianfar2023phononic}
\bibinfo{author}{Kianfar, A.} \& \bibinfo{author}{Hussein, M.~I.}
\newblock \bibinfo{journal}{\bibinfo{title}{Phononic-subsurface flow stabilization by subwavelength locally resonant metamaterials}}.
\newblock {\emph{\JournalTitle{New Journal of Physics}}} \textbf{\bibinfo{volume}{25}}, \bibinfo{pages}{053021} (\bibinfo{year}{2023}).

\bibitem{michelis2023attenuation}
\bibinfo{author}{Michelis, T.}, \bibinfo{author}{Putranto, A.} \& \bibinfo{author}{Kotsonis, M.}
\newblock \bibinfo{journal}{\bibinfo{title}{Attenuation of tollmien--schlichting waves using resonating surface-embedded phononic crystals}}.
\newblock {\emph{\JournalTitle{Physics of Fluids}}} \textbf{\bibinfo{volume}{35}} (\bibinfo{year}{2023}).

\bibitem{willey2023multi}
\bibinfo{author}{Willey, C.~L.} \emph{et~al.}
\newblock \bibinfo{journal}{\bibinfo{title}{Multi-input multi-output phononic subsurfaces for passive boundary layer transition delay}}.
\newblock {\emph{\JournalTitle{Journal of Fluids and Structures}}} \textbf{\bibinfo{volume}{121}}, \bibinfo{pages}{103936} (\bibinfo{year}{2023}).

\bibitem{schmidt2025perturbation}
\bibinfo{author}{Schmidt, R.}, \bibinfo{author}{Yousef, H.}, \bibinfo{author}{Roy, I.}, \bibinfo{author}{Scalo, C.} \& \bibinfo{author}{Nouh, M.}
\newblock \bibinfo{journal}{\bibinfo{title}{Perturbation energy extraction from a fluid via a subsurface acoustic diode with sustained downstream attenuation}}.
\newblock {\emph{\JournalTitle{Journal of Applied Physics}}} \textbf{\bibinfo{volume}{137}} (\bibinfo{year}{2025}).

\bibitem{navarro2025stabilization}
\bibinfo{author}{Navarro, J.~D.} \emph{et~al.}
\newblock \bibinfo{journal}{\bibinfo{title}{Stabilization of hypersonic shockwave/boundary-layer interactions with phononic metamaterials}}.
\newblock {\emph{\JournalTitle{Matter}}}  (\bibinfo{year}{2025}).

\bibitem{Michelis2023b}
\bibinfo{author}{Michelis, T.}, \bibinfo{author}{de~Koning, C.} \& \bibinfo{author}{Kotsonis, M.}
\newblock \bibinfo{journal}{\bibinfo{title}{On the interaction of tollmien–schlichting waves with a wall-embedded helmholtz resonator}}.
\newblock {\emph{\JournalTitle{Physics of Fluids}}} \textbf{\bibinfo{volume}{35}}, \bibinfo{pages}{034104}, \doiprefix\url{10.1063/5.0141685} (\bibinfo{year}{2023}).

\bibitem{dacome2024small}
\bibinfo{author}{Dacome, G.}, \bibinfo{author}{Siebols, R.} \& \bibinfo{author}{Baars, W.}
\newblock \bibinfo{journal}{\bibinfo{title}{Small-scale helmholtz resonators with grazing turbulent boundary layer flow}}.
\newblock {\emph{\JournalTitle{Journal of Turbulence}}} \textbf{\bibinfo{volume}{25}}, \bibinfo{pages}{461--481} (\bibinfo{year}{2024}).

\bibitem{hassanein2024interaction}
\bibinfo{author}{Hassanein, A.}, \bibinfo{author}{Modesti, D.}, \bibinfo{author}{Scarano, F.} \& \bibinfo{author}{Baars, W.~J.}
\newblock \bibinfo{journal}{\bibinfo{title}{Interaction of an inner-scaled helmholtz resonator with boundary-layer turbulence}}.
\newblock {\emph{\JournalTitle{Physical Review Fluids}}} \textbf{\bibinfo{volume}{9}}, \bibinfo{pages}{114610} (\bibinfo{year}{2024}).

\bibitem{benjamin1960effects}
\bibinfo{author}{Benjamin, T.~B.}
\newblock \bibinfo{journal}{\bibinfo{title}{Effects of a flexible boundary on hydrodynamic stability}}.
\newblock {\emph{\JournalTitle{Journal of Fluid Mechanics}}} \textbf{\bibinfo{volume}{9}}, \bibinfo{pages}{513--532} (\bibinfo{year}{1960}).

\bibitem{carpenter-et-al:ja2001}
\bibinfo{author}{Carpenter, P.~W.}, \bibinfo{author}{Lucey, A.~D.} \& \bibinfo{author}{Davies, C.}
\newblock \bibinfo{journal}{\bibinfo{title}{Progress on the use of compliant walls for laminar-flow control}}.
\newblock {\emph{\JournalTitle{Journal of Aircraft}}} \textbf{\bibinfo{volume}{38}}, \bibinfo{pages}{504--512}, \doiprefix\url{10.2514/2.2790} (\bibinfo{year}{2001}).

\bibitem{Fabbiane2025}
\bibinfo{author}{Fabbiane, N.}, \bibinfo{author}{Marquet, O.}, \bibinfo{author}{Pierpaoli, L.}, \bibinfo{author}{Cottereau, R.} \& \bibinfo{author}{Couliou, M.}
\newblock \bibinfo{journal}{\bibinfo{title}{Phononic compliant surfaces for the suppression of travelling wave flutter instabilities in boundary-layer flows}}.
\newblock {\emph{\JournalTitle{Journal of Fluid Mechanics}}}  (\bibinfo{year}{in press}).

\bibitem{pfister:phd2019}
\bibinfo{author}{Pfister, J.-L.}
\newblock \emph{\bibinfo{title}{Instabilities and optimization of elastic strutures interacting with laminar flows}}.
\newblock Ph.D. thesis, \bibinfo{school}{Universit\'e Paris-Saclay}, \bibinfo{address}{France} (\bibinfo{year}{2019}).

\bibitem{walsh1978drag}
\bibinfo{author}{Walsh, M.} \& \bibinfo{author}{Weinstein, L.}
\newblock \bibinfo{title}{Drag and heat transfer on surfaces with small longitudinal fins}.
\newblock In \emph{\bibinfo{booktitle}{11th Fluid and PlasmaDynamics Conference}}, \bibinfo{pages}{1161} (\bibinfo{year}{1978}).

\bibitem{pollard1998passive}
\bibinfo{author}{Pollard, A.}
\newblock \bibinfo{journal}{\bibinfo{title}{Passive and active control of near-wall turbulence}}.
\newblock {\emph{\JournalTitle{Progress in aerospace sciences}}} \textbf{\bibinfo{volume}{33}}, \bibinfo{pages}{689--708} (\bibinfo{year}{1998}).

\bibitem{ghaemi2020passive}
\bibinfo{author}{Ghaemi, S.}
\newblock \bibinfo{journal}{\bibinfo{title}{Passive and active control of turbulent flows}}.
\newblock {\emph{\JournalTitle{Physics of fluids}}} \textbf{\bibinfo{volume}{32}} (\bibinfo{year}{2020}).

\bibitem{sharma2020turbulent}
\bibinfo{author}{Sharma, A.} \& \bibinfo{author}{Garc{\'\i}a-Mayoral, R.}
\newblock \bibinfo{journal}{\bibinfo{title}{Turbulent flows over dense filament canopies}}.
\newblock {\emph{\JournalTitle{Journal of Fluid Mechanics}}} \textbf{\bibinfo{volume}{888}}, \bibinfo{pages}{A2} (\bibinfo{year}{2020}).

\bibitem{Zoppini2022}
\bibinfo{author}{Zoppini, G.}, \bibinfo{author}{Michelis, T.}, \bibinfo{author}{Ragni, D.} \& \bibinfo{author}{Kotsonis, M.}
\newblock \bibinfo{journal}{\bibinfo{title}{Cancellation of crossflow instabilities through multiple discrete roughness elements forcing}}.
\newblock {\emph{\JournalTitle{Phys. Rev. Fluids}}} \textbf{\bibinfo{volume}{7}}, \bibinfo{pages}{123902}, \doiprefix\url{10.1103/PhysRevFluids.7.123902} (\bibinfo{year}{2022}).

\bibitem{zoppini2023receptivity}
\bibinfo{author}{Zoppini, G.}
\newblock \emph{\bibinfo{title}{Receptivity of Swept Wing Boundary Layers to Surface Roughness: Diagnostics and extension to flow control}}.
\newblock Ph.D. thesis, \bibinfo{school}{Delft University of Technology} (\bibinfo{year}{2023}).

\bibitem{piest2024mitigation}
\bibinfo{author}{Piest, B.}, \bibinfo{author}{Druetta, P.} \& \bibinfo{author}{Krushynska, A.}
\newblock \bibinfo{journal}{\bibinfo{title}{Mitigation of flow-induced vibrations in high-speed flows using triply periodic minimal surface structures}}.
\newblock {\emph{\JournalTitle{Physics of Fluids}}} \textbf{\bibinfo{volume}{36}} (\bibinfo{year}{2024}).

\bibitem{hickey2020targeted}
\bibinfo{author}{Hickey, J.-P.}, \bibinfo{author}{Younes, K.}, \bibinfo{author}{Yao, M.~X.}, \bibinfo{author}{Fan, D.} \& \bibinfo{author}{Mouallem, J.}
\newblock \bibinfo{journal}{\bibinfo{title}{Targeted turbulent structure control in wall-bounded flows via localized heating}}.
\newblock {\emph{\JournalTitle{Physics of Fluids}}} \textbf{\bibinfo{volume}{32}} (\bibinfo{year}{2020}).

\bibitem{luo2003design}
\bibinfo{author}{Luo, H.} \& \bibinfo{author}{Bewley, T.~R.}
\newblock \bibinfo{title}{Design, modeling, and optimization of compliant tensegrity fabrics for the reduction of turbulent skin friction}.
\newblock In \emph{\bibinfo{booktitle}{Smart Structures and Materials 2003: Modeling, Signal Processing, and Control}}, vol. \bibinfo{volume}{5049}, \bibinfo{pages}{460--470} (\bibinfo{organization}{SPIE}, \bibinfo{year}{2003}).

\bibitem{wen2023dynamic}
\bibinfo{author}{Wen, X.} \emph{et~al.}
\newblock \bibinfo{journal}{\bibinfo{title}{Dynamic kirigami structures for wake flow control behind a circular cylinder}}.
\newblock {\emph{\JournalTitle{Physics of Fluids}}} \textbf{\bibinfo{volume}{35}} (\bibinfo{year}{2023}).

\bibitem{fedorov2001stabilization}
\bibinfo{author}{Fedorov, A.~V.}, \bibinfo{author}{Malmuth, N.~D.}, \bibinfo{author}{Rasheed, A.} \& \bibinfo{author}{Hornung, H.~G.}
\newblock \bibinfo{journal}{\bibinfo{title}{Stabilization of hypersonic boundary layers by porous coatings}}.
\newblock {\emph{\JournalTitle{AIAA journal}}} \textbf{\bibinfo{volume}{39}}, \bibinfo{pages}{605--610} (\bibinfo{year}{2001}).

\bibitem{rosti2018turbulent}
\bibinfo{author}{Rosti, M.~E.}, \bibinfo{author}{Brandt, L.} \& \bibinfo{author}{Pinelli, A.}
\newblock \bibinfo{journal}{\bibinfo{title}{Turbulent channel flow over an anisotropic porous wall--drag increase and reduction}}.
\newblock {\emph{\JournalTitle{Journal of Fluid Mechanics}}} \textbf{\bibinfo{volume}{842}}, \bibinfo{pages}{381--394} (\bibinfo{year}{2018}).

\bibitem{Klausmann}
\bibinfo{author}{Klausmann, K.} \& \bibinfo{author}{Ruck, B.}
\newblock \bibinfo{journal}{\bibinfo{title}{Drag reduction of circular cylinders by porous coating on the leeward side}}.
\newblock {\emph{\JournalTitle{Journal of Fluid Mechanics}}} \textbf{\bibinfo{volume}{813}}, \doiprefix\url{10.1017/jfm.2016.757} (\bibinfo{year}{2017}).

\bibitem{kachanov:arfm1994}
\bibinfo{author}{Kachanov, Y.~S.}
\newblock \bibinfo{journal}{\bibinfo{title}{Physical mechanisms of laminar-boundary-layer transition}}.
\newblock {\emph{\JournalTitle{Annual Review of Fluid Mechanics}}} \textbf{\bibinfo{volume}{26}}, \bibinfo{pages}{411--482}, \doiprefix\url{10.1146/annurev.fl.26.010194.002211} (\bibinfo{year}{1994}).

\bibitem{milling1981tollmien}
\bibinfo{author}{Milling, R.~W.}
\newblock \bibinfo{journal}{\bibinfo{title}{Tollmien--schlichting wave cancellation}}.
\newblock {\emph{\JournalTitle{Physics of Fluids}}} \textbf{\bibinfo{volume}{24}}, \bibinfo{pages}{979--981} (\bibinfo{year}{1981}).

\bibitem{hussein2014dynamics}
\bibinfo{author}{Hussein, M.~I.}, \bibinfo{author}{Leamy, M.~J.} \& \bibinfo{author}{Ruzzene, M.}
\newblock \bibinfo{journal}{\bibinfo{title}{Dynamics of phononic materials and structures: Historical origins, recent progress, and future outlook}}.
\newblock {\emph{\JournalTitle{Applied Mechanics Reviews}}} \textbf{\bibinfo{volume}{66}} (\bibinfo{year}{2014}).

\bibitem{Barnes_2021}
\bibinfo{author}{Barnes, C.~J.}, \bibinfo{author}{Willey, C.~L.}, \bibinfo{author}{Rosenberg, K.}, \bibinfo{author}{Medina, A.} \& \bibinfo{author}{Juhl, A.~T.}
\newblock \bibinfo{title}{Initial computational investigation toward passive transition delay using a phononic subsurface}.
\newblock In \emph{\bibinfo{booktitle}{AIAA Scitech 2021 Forum}}, \bibinfo{pages}{1454} (\bibinfo{year}{2021}).

\bibitem{kianfar2023local}
\bibinfo{author}{Kianfar, A.} \& \bibinfo{author}{Hussein, M.~I.}
\newblock \bibinfo{journal}{\bibinfo{title}{Local flow control by phononic subsurfaces over extended spatial domains}}.
\newblock {\emph{\JournalTitle{Journal of Applied Physics}}} \textbf{\bibinfo{volume}{134}} (\bibinfo{year}{2023}).

\bibitem{hussein2025scatterless}
\bibinfo{author}{Hussein, M.~I.}, \bibinfo{author}{Roca, D.}, \bibinfo{author}{Harris, A.~R.} \& \bibinfo{author}{Kianfar, A.}
\newblock \bibinfo{journal}{\bibinfo{title}{Scatterless interferences: Delay of laminar-to-turbulent flow transition by a lattice of subsurface phonons}}.
\newblock {\emph{\JournalTitle{arXiv preprint arXiv:2503.18835}}}  (\bibinfo{year}{2025}).

\bibitem{al2023theory}
\bibinfo{author}{Al~Ba'ba'a, H.~B.}, \bibinfo{author}{Willey, C.~L.}, \bibinfo{author}{Chen, V.~W.}, \bibinfo{author}{Juhl, A.~T.} \& \bibinfo{author}{Nouh, M.}
\newblock \bibinfo{journal}{\bibinfo{title}{Theory of truncation resonances in continuum rod-based phononic crystals with generally asymmetric unit cells}}.
\newblock {\emph{\JournalTitle{Advanced Theory and Simulations}}} \textbf{\bibinfo{volume}{6}}, \bibinfo{pages}{2200700} (\bibinfo{year}{2023}).

\bibitem{rosa2023material}
\bibinfo{author}{Rosa, M.~I.}, \bibinfo{author}{Davis, B.~L.}, \bibinfo{author}{Liu, L.}, \bibinfo{author}{Ruzzene, M.} \& \bibinfo{author}{Hussein, M.~I.}
\newblock \bibinfo{journal}{\bibinfo{title}{Material vs. structure: Topological origins of band-gap truncation resonances in periodic structures}}.
\newblock {\emph{\JournalTitle{Physical Review Materials}}} \textbf{\bibinfo{volume}{7}}, \bibinfo{pages}{124201} (\bibinfo{year}{2023}).

\bibitem{benjamin:jfm1960}
\bibinfo{author}{Benjamin, T.~B.}
\newblock \bibinfo{journal}{\bibinfo{title}{Effects of a flexible boundary on hydrodynamic stability}}.
\newblock {\emph{\JournalTitle{Journal of Fluid Mechanics}}} \textbf{\bibinfo{volume}{9}}, \bibinfo{pages}{513--532}, \doiprefix\url{10.1017/S0022112060001286} (\bibinfo{year}{1960}).

\bibitem{landahl:jfm1962}
\bibinfo{author}{Landahl, M.~T.}
\newblock \bibinfo{journal}{\bibinfo{title}{On the stability of a laminar incompressible boundary layer over a flexible surface}}.
\newblock {\emph{\JournalTitle{Journal of Fluid Mechanics}}} \textbf{\bibinfo{volume}{13}}, \bibinfo{pages}{609–632}, \doiprefix\url{10.1017/S002211206200097X} (\bibinfo{year}{1962}).

\bibitem{benjamin:jfm1963}
\bibinfo{author}{Benjamin, T.~B.}
\newblock \bibinfo{journal}{\bibinfo{title}{The threefold classification of unstable disturbances in flexible surfaces bounding inviscid flows}}.
\newblock {\emph{\JournalTitle{Journal of Fluid Mechanics}}} \textbf{\bibinfo{volume}{16}}, \bibinfo{pages}{436–450}, \doiprefix\url{10.1017/S0022112063000884} (\bibinfo{year}{1963}).

\bibitem{carpenter-garrad:jfm1985}
\bibinfo{author}{Carpenter, P.~W.} \& \bibinfo{author}{Garrad, A.~D.}
\newblock \bibinfo{journal}{\bibinfo{title}{The hydrodynamic stability of flow over {Kramer}-type compliant surfaces. {Part} 1. {Tollmien}-{Schlichting} instabilities}}.
\newblock {\emph{\JournalTitle{Journal of Fluid Mechanics}}} \textbf{\bibinfo{volume}{155}}, \bibinfo{pages}{465--510}, \doiprefix\url{10.1017/S0022112085001902} (\bibinfo{year}{1985}).

\bibitem{carpenter-garrad:jfm1986}
\bibinfo{author}{Carpenter, P.~W.} \& \bibinfo{author}{Garrad, A.~D.}
\newblock \bibinfo{journal}{\bibinfo{title}{The hydrodynamic stability of flow over kramer-type compliant surfaces. part 2. flow-induced surface instabilities}}.
\newblock {\emph{\JournalTitle{Journal of Fluid Mechanics}}} \textbf{\bibinfo{volume}{170}}, \bibinfo{pages}{199–232}, \doiprefix\url{10.1017/S002211208600085X} (\bibinfo{year}{1986}).

\bibitem{yeo:jfm1988}
\bibinfo{author}{Yeo, K.~S.}
\newblock \bibinfo{journal}{\bibinfo{title}{The stability of boundary-layer flow over single-and multi-layer viscoelastic walls}}.
\newblock {\emph{\JournalTitle{Journal of Fluid Mechanics}}} \textbf{\bibinfo{volume}{196}}, \bibinfo{pages}{359--408}, \doiprefix\url{10.1017/S0022112088002745} (\bibinfo{year}{1988}).

\bibitem{tsigklifis-lucey:jfm2017}
\bibinfo{author}{Tsigklifis, K.} \& \bibinfo{author}{Lucey, A.~D.}
\newblock \bibinfo{journal}{\bibinfo{title}{The interaction of {Blasius} boundary-layer flow with a compliant panel: global, local and transient analyses}}.
\newblock {\emph{\JournalTitle{Journal of Fluid Mechanics}}} \textbf{\bibinfo{volume}{827}}, \bibinfo{pages}{155--193}, \doiprefix\url{10.1017/jfm.2017.453} (\bibinfo{year}{2017}).

\bibitem{pfister-et-al:jfm2022}
\bibinfo{author}{Pfister, J.-L.}, \bibinfo{author}{Fabbiane, N.} \& \bibinfo{author}{Marquet, O.}
\newblock \bibinfo{journal}{\bibinfo{title}{{Global stability and resolvent analyses of laminar boundary-layer flow interacting with visco-elastic patches}}}.
\newblock {\emph{\JournalTitle{Journal of Fluid Mechanics}}} \textbf{\bibinfo{volume}{937}}, \bibinfo{pages}{A1}, \doiprefix\url{10.1017/jfm.2022.72} (\bibinfo{year}{2022}).

\bibitem{Zhao2019}
\bibinfo{author}{Zhao, R.}, \bibinfo{author}{Liu, T.}, \bibinfo{author}{Wen, C.-y.}, \bibinfo{author}{Zhu, J.} \& \bibinfo{author}{Cheng, L.}
\newblock \bibinfo{journal}{\bibinfo{title}{Impedance-near-zero acoustic metasurface for hypersonic boundary-layer flow stabilization}}.
\newblock {\emph{\JournalTitle{Phys. Rev. Appl.}}} \textbf{\bibinfo{volume}{11}}, \bibinfo{pages}{044015}, \doiprefix\url{10.1103/PhysRevApplied.11.044015} (\bibinfo{year}{2019}).

\bibitem{Saric2003}
\bibinfo{author}{Saric, W.~S.}, \bibinfo{author}{Reed, H.~L.} \& \bibinfo{author}{White, E.~B.}
\newblock \bibinfo{journal}{\bibinfo{title}{Stability and transition of three-dimensional boundary layers}}.
\newblock {\emph{\JournalTitle{Annual Review of Fluid Mechanics}}} \textbf{\bibinfo{volume}{35}}, \bibinfo{pages}{413--440}, \doiprefix\url{https://doi.org/10.1146/annurev.fluid.35.101101.161045} (\bibinfo{year}{2003}).

\bibitem{ivanov2019delay}
\bibinfo{author}{Ivanov, A.} \& \bibinfo{author}{Mischenko, D.}
\newblock \bibinfo{title}{Delay of laminar-turbulent transition on swept-wing with help of sweeping surface relief}.
\newblock In \emph{\bibinfo{booktitle}{AIP Conference Proceedings}}, vol. \bibinfo{volume}{2125} (\bibinfo{organization}{AIP Publishing}, \bibinfo{year}{2019}).

\bibitem{ustinov2018cross}
\bibinfo{author}{Ustinov, M.} \& \bibinfo{author}{Ivanov, A.}
\newblock \bibinfo{title}{Cross-flow dominated transition control by surface micro-relief}.
\newblock In \emph{\bibinfo{booktitle}{AIP Conference Proceedings}}, vol. \bibinfo{volume}{2027} (\bibinfo{organization}{AIP Publishing}, \bibinfo{year}{2018}).

\bibitem{gad1991separation}
\bibinfo{author}{Gad-el Hak, M.} \& \bibinfo{author}{Bushnell, D.~M.}
\newblock \bibinfo{journal}{\bibinfo{title}{Separation control}}.
\newblock {\emph{\JournalTitle{Journal of fluids engineering}}} \textbf{\bibinfo{volume}{113}}, \bibinfo{pages}{5--30} (\bibinfo{year}{1991}).

\bibitem{ashill2005review}
\bibinfo{author}{Ashill, P.}, \bibinfo{author}{Fulker, J.} \& \bibinfo{author}{Hackett, K.}
\newblock \bibinfo{journal}{\bibinfo{title}{A review of recent developments in flow control}}.
\newblock {\emph{\JournalTitle{The Aeronautical Journal}}} \textbf{\bibinfo{volume}{109}}, \bibinfo{pages}{205--232} (\bibinfo{year}{2005}).

\bibitem{Cattafesta}
\bibinfo{author}{Cattafesta, L.~N.} \& \bibinfo{author}{Sheplak, M.}
\newblock \bibinfo{journal}{\bibinfo{title}{Actuators for active flow control}}.
\newblock {\emph{\JournalTitle{Annual Review of Fluid Mechanics}}} \textbf{\bibinfo{volume}{43}}, \doiprefix\url{10.1146/annurev-fluid-122109-160634} (\bibinfo{year}{2011}).

\bibitem{lin2002review}
\bibinfo{author}{Lin, J.~C.}
\newblock \bibinfo{journal}{\bibinfo{title}{Review of research on low-profile vortex generators to control boundary-layer separation}}.
\newblock {\emph{\JournalTitle{Progress in aerospace sciences}}} \textbf{\bibinfo{volume}{38}}, \bibinfo{pages}{389--420} (\bibinfo{year}{2002}).

\bibitem{Corke2010}
\bibinfo{author}{Corke, T.~C.}, \bibinfo{author}{Enloe, C.~L.} \& \bibinfo{author}{Wilkinson, S.~P.}
\newblock \bibinfo{journal}{\bibinfo{title}{Dielectric barrier discharge plasma actuators for flow control}}.
\newblock {\emph{\JournalTitle{Annual Review of Fluid Mechanics}}} \textbf{\bibinfo{volume}{42}}, \doiprefix\url{10.1146/annurev-fluid-121108-145550} (\bibinfo{year}{2010}).

\bibitem{Glezer2002}
\bibinfo{author}{Glezer, A.} \& \bibinfo{author}{Amitay, M.}
\newblock \bibinfo{journal}{\bibinfo{title}{Synthetic jets}}.
\newblock {\emph{\JournalTitle{Annual Review of Fluid Mechanics}}} \textbf{\bibinfo{volume}{34}}, \doiprefix\url{10.1146/annurev.fluid.34.090501.094913} (\bibinfo{year}{2002}).

\bibitem{Seol2023}
\bibinfo{author}{Seol, C.}, \bibinfo{author}{Hong, J.} \& \bibinfo{author}{Kim, T.}
\newblock \bibinfo{journal}{\bibinfo{title}{Flow around porous square cylinders with a periodic and scalable structure}}.
\newblock {\emph{\JournalTitle{Experimental Thermal and Fluid Science}}} \textbf{\bibinfo{volume}{144}}, \doiprefix\url{10.1016/j.expthermflusci.2023.110864} (\bibinfo{year}{2023}).

\bibitem{Seol2024}
\bibinfo{author}{Seol, C.}, \bibinfo{author}{Kim, T.} \& \bibinfo{author}{Kim, T.}
\newblock \bibinfo{journal}{\bibinfo{title}{The effect of permeability on the flow structure of porous square cylinders}}.
\newblock {\emph{\JournalTitle{Journal of Fluid Mechanics}}} \textbf{\bibinfo{volume}{985}}, \doiprefix\url{10.1017/jfm.2024.311} (\bibinfo{year}{2024}).

\bibitem{visbal2023passive}
\bibinfo{author}{Visbal, M.~R.} \& \bibinfo{author}{Garmann, D.~J.}
\newblock \bibinfo{journal}{\bibinfo{title}{Passive control of dynamic stall using a flow-driven micro-cavity actuator}}.
\newblock {\emph{\JournalTitle{Theoretical and Computational Fluid Dynamics}}} \textbf{\bibinfo{volume}{37}}, \bibinfo{pages}{289--303} (\bibinfo{year}{2023}).

\bibitem{gad1996modern}
\bibinfo{author}{Gad-el Hak, M.}
\newblock \bibinfo{journal}{\bibinfo{title}{Modern developments in flow control}}.
\newblock {\emph{\JournalTitle{Appl. Mech. Rev.}}} \textbf{\bibinfo{volume}{49}}, \bibinfo{pages}{365--379} (\bibinfo{year}{1996}).

\bibitem{abbas2017drag}
\bibinfo{author}{Abbas, A.} \emph{et~al.}
\newblock \bibinfo{journal}{\bibinfo{title}{Drag reduction via turbulent boundary layer flow control}}.
\newblock {\emph{\JournalTitle{Science China Technological Sciences}}} \textbf{\bibinfo{volume}{60}}, \bibinfo{pages}{1281--1290} (\bibinfo{year}{2017}).

\bibitem{corke2018active}
\bibinfo{author}{Corke, T.~C.} \& \bibinfo{author}{Thomas, F.~O.}
\newblock \bibinfo{journal}{\bibinfo{title}{Active and passive turbulent boundary-layer drag reduction}}.
\newblock {\emph{\JournalTitle{AIAA journal}}} \textbf{\bibinfo{volume}{56}}, \bibinfo{pages}{3835--3847} (\bibinfo{year}{2018}).

\bibitem{warsop2001current}
\bibinfo{author}{Warsop, C.}
\newblock \bibinfo{title}{Current status and prospects for turbulent flow control}.
\newblock In \emph{\bibinfo{booktitle}{Aerodynamic Drag Reduction Technologies: Proceedings of the CEAS/DragNet European Drag Reduction Conference, 19--21 June 2000, Potsdam, Germany}}, \bibinfo{pages}{269--277} (\bibinfo{organization}{Springer}, \bibinfo{year}{2001}).

\bibitem{hehner2019stokes}
\bibinfo{author}{Hehner, M.~T.}, \bibinfo{author}{Gatti, D.} \& \bibinfo{author}{Kriegseis, J.}
\newblock \bibinfo{journal}{\bibinfo{title}{Stokes-layer formation under absence of moving parts—a novel oscillatory plasma actuator design for turbulent drag reduction}}.
\newblock {\emph{\JournalTitle{Physics of Fluids}}} \textbf{\bibinfo{volume}{31}} (\bibinfo{year}{2019}).

\bibitem{lin2024control}
\bibinfo{author}{Lin, C.-T.}, \bibinfo{author}{Ramakrishnan, V.}, \bibinfo{author}{Goza, A.}, \bibinfo{author}{Matlack, K.} \& \bibinfo{author}{Bae, J.}
\newblock \bibinfo{title}{Control for turbulent drag reduction by wall-normal blowing and suction}.
\newblock In \emph{\bibinfo{booktitle}{APS Division of Fluid Dynamics Meeting Abstracts}}, \bibinfo{pages}{L33--002} (\bibinfo{year}{2024}).

\bibitem{Bobinski2018BackscatteringChannel}
\bibinfo{author}{Bobinski, T.}, \bibinfo{author}{Maurel, A.}, \bibinfo{author}{Petitjeans, P.} \& \bibinfo{author}{Pagneux, V.}
\newblock \bibinfo{journal}{\bibinfo{title}{{Backscattering reduction for resonating obstacle in water-wave channel}}}.
\newblock {\emph{\JournalTitle{Journal of Fluid Mechanics}}} \textbf{\bibinfo{volume}{845}}, \doiprefix\url{10.1017/jfm.2018.302} (\bibinfo{year}{2018}).

\bibitem{Bobinski2015ExperimentalFocusing}
\bibinfo{author}{Bobinski, T.}, \bibinfo{author}{Eddi, A.}, \bibinfo{author}{Petitjeans, P.}, \bibinfo{author}{Maurel, A.} \& \bibinfo{author}{Pagneux, V.}
\newblock \bibinfo{journal}{\bibinfo{title}{{Experimental demonstration of epsilon-near-zero water waves focusing}}}.
\newblock {\emph{\JournalTitle{Applied Physics Letters}}} \textbf{\bibinfo{volume}{107}}, \doiprefix\url{10.1063/1.4926362} (\bibinfo{year}{2015}).

\bibitem{Sharma2020OnMaterial}
\bibinfo{author}{Sharma, G.~S.}, \bibinfo{author}{Skvortsov, A.}, \bibinfo{author}{MacGillivray, I.} \& \bibinfo{author}{Kessissoglou, N.}
\newblock \bibinfo{journal}{\bibinfo{title}{{On superscattering of sound waves by a lattice of disk-shaped cavities in a soft material}}}.
\newblock {\emph{\JournalTitle{Applied Physics Letters}}} \textbf{\bibinfo{volume}{116}}, \doiprefix\url{10.1063/1.5130695} (\bibinfo{year}{2020}).

\bibitem{Berraquero2013ExperimentalShifter}
\bibinfo{author}{Berraquero, C.~P.}, \bibinfo{author}{Maurel, A.}, \bibinfo{author}{Petitjeans, P.} \& \bibinfo{author}{Pagneux, V.}
\newblock \bibinfo{journal}{\bibinfo{title}{{Experimental realization of a water-wave metamaterial shifter}}}.
\newblock {\emph{\JournalTitle{Physical Review E - Statistical, Nonlinear, and Soft Matter Physics}}} \textbf{\bibinfo{volume}{88}}, \doiprefix\url{10.1103/PhysRevE.88.051002} (\bibinfo{year}{2013}).

\bibitem{Hu2011NegativeArrays}
\bibinfo{author}{Hu, X.}, \bibinfo{author}{Chan, C.~T.}, \bibinfo{author}{Ho, K.~M.} \& \bibinfo{author}{Zi, J.}
\newblock \bibinfo{journal}{\bibinfo{title}{{Negative effective gravity in water waves by periodic resonator arrays}}}.
\newblock {\emph{\JournalTitle{Physical Review Letters}}} \textbf{\bibinfo{volume}{106}}, \doiprefix\url{10.1103/PhysRevLett.106.174501} (\bibinfo{year}{2011}).

\bibitem{Hu2013ExperimentalWaves}
\bibinfo{author}{Hu, X.}, \bibinfo{author}{Yang, J.}, \bibinfo{author}{Zi, J.}, \bibinfo{author}{Chan, C.~T.} \& \bibinfo{author}{Ho, K.~M.}
\newblock \bibinfo{journal}{\bibinfo{title}{{Experimental observation of negative effective gravity in water waves}}}.
\newblock {\emph{\JournalTitle{Scientific Reports}}} \textbf{\bibinfo{volume}{3}}, \doiprefix\url{10.1038/srep01916} (\bibinfo{year}{2013}).

\bibitem{Hu2003CompleteBottom}
\bibinfo{author}{Hu, X.}, \bibinfo{author}{Shen, Y.}, \bibinfo{author}{Liu, X.}, \bibinfo{author}{Fu, R.} \& \bibinfo{author}{Zi, J.}
\newblock \bibinfo{journal}{\bibinfo{title}{{Complete band gaps for liquid surface waves propagating over a periodically drilled bottom}}}.
\newblock {\emph{\JournalTitle{Physical Review E - Statistical Physics, Plasmas, Fluids, and Related Interdisciplinary Topics}}} \textbf{\bibinfo{volume}{68}}, \doiprefix\url{10.1103/PhysRevE.68.066308} (\bibinfo{year}{2003}).

\bibitem{Kar2020BraggTrenches}
\bibinfo{author}{Kar, P.}, \bibinfo{author}{Sahoo, T.} \& \bibinfo{author}{Meylan, M.~H.}
\newblock \bibinfo{journal}{\bibinfo{title}{{Bragg scattering of long waves by an array of floating flexible plates in the presence of multiple submerged trenches}}}.
\newblock {\emph{\JournalTitle{Physics of Fluids}}} \textbf{\bibinfo{volume}{32}}, \doiprefix\url{10.1063/5.0017930} (\bibinfo{year}{2020}).

\bibitem{Hu2005RefractionArrays}
\bibinfo{author}{Hu, X.} \& \bibinfo{author}{Chan, C.~T.}
\newblock \bibinfo{journal}{\bibinfo{title}{{Refraction of water waves by periodic cylinder arrays}}}.
\newblock {\emph{\JournalTitle{Physical Review Letters}}} \textbf{\bibinfo{volume}{95}}, \doiprefix\url{10.1103/PhysRevLett.95.154501} (\bibinfo{year}{2005}).

\bibitem{Bennetts2018GradedWaves}
\bibinfo{author}{Bennetts, L.~G.}, \bibinfo{author}{Peter, M.~A.} \& \bibinfo{author}{Craster, R.~V.}
\newblock \bibinfo{journal}{\bibinfo{title}{{Graded resonator arrays for spatial frequency separation and amplification of water waves}}}.
\newblock {\emph{\JournalTitle{Journal of Fluid Mechanics}}} \textbf{\bibinfo{volume}{854}}, \doiprefix\url{10.1017/jfm.2018.648} (\bibinfo{year}{2018}).

\bibitem{Wilks2022RainbowBarriers}
\bibinfo{author}{Wilks, B.}, \bibinfo{author}{Montiel, F.} \& \bibinfo{author}{Wakes, S.}
\newblock \bibinfo{journal}{\bibinfo{title}{{Rainbow reflection and broadband energy absorption of water waves by graded arrays of vertical barriers}}}.
\newblock {\emph{\JournalTitle{Journal of Fluid Mechanics}}} \textbf{\bibinfo{volume}{941}}, \doiprefix\url{10.1017/jfm.2022.302} (\bibinfo{year}{2022}).

\bibitem{Archer2020ExperimentalArrays}
\bibinfo{author}{Archer, A.~J.} \emph{et~al.}
\newblock \bibinfo{journal}{\bibinfo{title}{{Experimental realization of broadband control of water-wave-energy amplification in chirped arrays}}}.
\newblock {\emph{\JournalTitle{Physical Review Fluids}}} \textbf{\bibinfo{volume}{5}}, \doiprefix\url{10.1103/PhysRevFluids.5.062801} (\bibinfo{year}{2020}).

\bibitem{DeVita2021AttenuatingMetamaterials}
\bibinfo{author}{De~Vita, F.}, \bibinfo{author}{De~Lillo, F.}, \bibinfo{author}{Bosia, F.} \& \bibinfo{author}{Onorato, M.}
\newblock \bibinfo{journal}{\bibinfo{title}{{Attenuating surface gravity waves with mechanical metamaterials}}}.
\newblock {\emph{\JournalTitle{Physics of Fluids}}} \textbf{\bibinfo{volume}{33}}, \doiprefix\url{10.1063/5.0048613} (\bibinfo{year}{2021}).

\bibitem{DeVita2021AWaves}
\bibinfo{author}{De~Vita, F.}, \bibinfo{author}{De~Lillo, F.}, \bibinfo{author}{Verzicco, R.} \& \bibinfo{author}{Onorato, M.}
\newblock \bibinfo{journal}{\bibinfo{title}{{A fully Eulerian solver for the simulation of multiphase flows with solid bodies: Application to surface gravity waves}}}.
\newblock {\emph{\JournalTitle{Journal of Computational Physics}}} \textbf{\bibinfo{volume}{438}}, \doiprefix\url{10.1016/j.jcp.2021.110355} (\bibinfo{year}{2021}).

\bibitem{qu2018three}
\bibinfo{author}{Qu, J.}, \bibinfo{author}{Kadic, M.} \& \bibinfo{author}{Wegener, M.}
\newblock \bibinfo{journal}{\bibinfo{title}{Three-dimensional poroelastic metamaterials with extremely negative or positive effective static volume compressibility}}.
\newblock {\emph{\JournalTitle{Extreme Mechanics Letters}}} \textbf{\bibinfo{volume}{22}}, \bibinfo{pages}{165--171} (\bibinfo{year}{2018}).

\bibitem{liu2000locally}
\bibinfo{author}{Liu, Z.} \emph{et~al.}
\newblock \bibinfo{journal}{\bibinfo{title}{Locally resonant sonic materials}}.
\newblock {\emph{\JournalTitle{science}}} \textbf{\bibinfo{volume}{289}}, \bibinfo{pages}{1734--1736} (\bibinfo{year}{2000}).

\bibitem{van2020turbulent}
\bibinfo{author}{Van~Buren, T.}, \bibinfo{author}{Floryan, D.}, \bibinfo{author}{Ding, L.}, \bibinfo{author}{Hellstr{\"o}m, L.} \& \bibinfo{author}{Smits, A.}
\newblock \bibinfo{journal}{\bibinfo{title}{Turbulent pipe flow response to a step change in surface roughness}}.
\newblock {\emph{\JournalTitle{Journal of Fluid Mechanics}}} \textbf{\bibinfo{volume}{904}}, \bibinfo{pages}{A38} (\bibinfo{year}{2020}).

\bibitem{liang2018wavefront}
\bibinfo{author}{Liang, B.}, \bibinfo{author}{Cheng, J.-c.} \& \bibinfo{author}{Qiu, C.-W.}
\newblock \bibinfo{journal}{\bibinfo{title}{Wavefront manipulation by acoustic metasurfaces: from physics and applications}}.
\newblock {\emph{\JournalTitle{Nanophotonics}}} \textbf{\bibinfo{volume}{7}}, \bibinfo{pages}{1191--1205} (\bibinfo{year}{2018}).

\bibitem{cheng2015ultra}
\bibinfo{author}{Cheng, Y.} \emph{et~al.}
\newblock \bibinfo{journal}{\bibinfo{title}{Ultra-sparse metasurface for high reflection of low-frequency sound based on artificial mie resonances}}.
\newblock {\emph{\JournalTitle{Nature materials}}} \textbf{\bibinfo{volume}{14}}, \bibinfo{pages}{1013--1019} (\bibinfo{year}{2015}).

\bibitem{frenzel2013three}
\bibinfo{author}{Frenzel, T.} \emph{et~al.}
\newblock \bibinfo{journal}{\bibinfo{title}{Three-dimensional labyrinthine acoustic metamaterials}}.
\newblock {\emph{\JournalTitle{Applied Physics Letters}}} \textbf{\bibinfo{volume}{103}} (\bibinfo{year}{2013}).

\bibitem{Palma2018}
\bibinfo{author}{Palma, G.}, \bibinfo{author}{Mao, H.}, \bibinfo{author}{Burghignoli, L.}, \bibinfo{author}{Göransson, P.} \& \bibinfo{author}{Iemma, U.}
\newblock \bibinfo{journal}{\bibinfo{title}{Acoustic metamaterials in aeronautics}}.
\newblock {\emph{\JournalTitle{Applied Sciences}}} \textbf{\bibinfo{volume}{8}}, \bibinfo{pages}{971} (\bibinfo{year}{2018}).

\bibitem{Geyer2010}
\bibinfo{author}{Geyer, T.~F.}, \bibinfo{author}{Sarradj, E.} \& \bibinfo{author}{Fritzsche, C.}
\newblock \bibinfo{journal}{\bibinfo{title}{Porous airfoils: noise reduction and boundary layer effects}}.
\newblock {\emph{\JournalTitle{International Journal of Aeroacoustics}}} \textbf{\bibinfo{volume}{9}}, \bibinfo{pages}{787--820} (\bibinfo{year}{2010}).

\bibitem{RubioCarpio2019}
\bibinfo{author}{Rubio~Carpio, A.} \emph{et~al.}
\newblock \bibinfo{journal}{\bibinfo{title}{Experimental characterization of the turbulent boundary layer over a porous trailing edge for noise abatement}}.
\newblock {\emph{\JournalTitle{Journal of Sound and Vibration}}} \textbf{\bibinfo{volume}{443}}, \bibinfo{pages}{537--558} (\bibinfo{year}{2019}).

\bibitem{Teruna2020}
\bibinfo{author}{Teruna, C.} \emph{et~al.}
\newblock \bibinfo{journal}{\bibinfo{title}{Noise reduction mechanisms of an open-cell metal-foam trailing edge}}.
\newblock {\emph{\JournalTitle{Journal of Fluid Mechanics}}} \textbf{\bibinfo{volume}{898}}, \bibinfo{pages}{A18} (\bibinfo{year}{2020}).

\bibitem{teruna2021}
\bibinfo{author}{Teruna, C.}, \bibinfo{author}{Avallone, F.}, \bibinfo{author}{Casalino, D.} \& \bibinfo{author}{Ragni, D.}
\newblock \bibinfo{journal}{\bibinfo{title}{Numerical investigation of leading edge noise reduction on a rod-airfoil configuration using porous materials and serrations}}.
\newblock {\emph{\JournalTitle{Journal of Sound and Vibration}}} \textbf{\bibinfo{volume}{494}}, \bibinfo{pages}{115880} (\bibinfo{year}{2021}).

\bibitem{Jawahar2023}
\bibinfo{author}{Jawahar, H.}, \bibinfo{author}{Karabasov, S.~A.} \& \bibinfo{author}{Azarpeyvand, M.}
\newblock \bibinfo{journal}{\bibinfo{title}{Jet installation noise reduction using porous treatments}}.
\newblock {\emph{\JournalTitle{Journal of Sound and Vibration}}} \textbf{\bibinfo{volume}{545}} (\bibinfo{year}{2023}).

\bibitem{Avallone2018}
\bibinfo{author}{Avallone, F.}, \bibinfo{author}{Ragni, D.} \& \bibinfo{author}{Casalino, D.}
\newblock \bibinfo{journal}{\bibinfo{title}{Impingement of a propeller-slipstream on a leading edge with a flow-permeable insert: A computational aeroacoustic study}}.
\newblock {\emph{\JournalTitle{International Journal of Aeroacoustics}}} \textbf{\bibinfo{volume}{17}}, \bibinfo{pages}{687--711} (\bibinfo{year}{2018}).

\bibitem{Zamponi2025}
\bibinfo{author}{Zamponi, R.}, \bibinfo{author}{Rubio~Carpio, A.}, \bibinfo{author}{Avallone, F.} \& \bibinfo{author}{Ragni, D.}
\newblock \bibinfo{journal}{\bibinfo{title}{Noise source analysis of porous fairings in a scaled landing gear model}}.
\newblock {\emph{\JournalTitle{Aerospace Science and Technology}}} \textbf{\bibinfo{volume}{158}}, \bibinfo{pages}{109885} (\bibinfo{year}{2025}).

\bibitem{Ross2024}
\bibinfo{author}{Ross, E.~P.}, \bibinfo{author}{Figueroa-Ibrahim, K.~M.}, \bibinfo{author}{Morris, S.~C.}, \bibinfo{author}{L., S.~D.} \& \bibinfo{author}{Bennett, G.~J.}
\newblock \bibinfo{journal}{\bibinfo{title}{Evaluating an additive manufactured acoustic metamaterial using the advanced noise control fan}}.
\newblock {\emph{\JournalTitle{AIAA Journal}}} \textbf{\bibinfo{volume}{62}}, \bibinfo{pages}{2783--2799} (\bibinfo{year}{2024}).

\bibitem{Tek1957}
\bibinfo{author}{Tek, M.}
\newblock \bibinfo{journal}{\bibinfo{title}{Development of a generalized {D}arcy equation}}.
\newblock {\emph{\JournalTitle{Journal of Petroleum Technology}}} \textbf{\bibinfo{volume}{9}}, \bibinfo{pages}{45--47} (\bibinfo{year}{1957}).

\bibitem{Luesutthiviboon2021}
\bibinfo{author}{Luesutthiviboon, S.}, \bibinfo{author}{Ragni, D.}, \bibinfo{author}{Avallone, F.} \& \bibinfo{author}{Snellen, M.}
\newblock \bibinfo{journal}{\bibinfo{title}{An alternative permeable topology design space for trailing-edge noise attenuation}}.
\newblock {\emph{\JournalTitle{International Journal of Aeroacoustics}}} \textbf{\bibinfo{volume}{20}}, \bibinfo{pages}{221--253} (\bibinfo{year}{2021}).

\bibitem{Johnson1987}
\bibinfo{author}{Johnson, D.~L.}, \bibinfo{author}{Koplik, J.} \& \bibinfo{author}{Dashen, R.}
\newblock \bibinfo{journal}{\bibinfo{title}{Theory of dynamic permeability and tortuosity in fluid-saturated porous media}}.
\newblock {\emph{\JournalTitle{Journal of Fluid Mechanics}}} \textbf{\bibinfo{volume}{176}}, \bibinfo{pages}{379--402} (\bibinfo{year}{1987}).

\bibitem{iemma2020}
\bibinfo{author}{Iemma, U.} \& \bibinfo{author}{Palma, G.}
\newblock \bibinfo{journal}{\bibinfo{title}{Design of metacontinua in the aeroacoustic spacetime}}.
\newblock {\emph{\JournalTitle{Scientific Reports}}} \textbf{\bibinfo{volume}{10}} (\bibinfo{year}{2020}).

\bibitem{zamponi2024}
\bibinfo{author}{Zamponi, R.}, \bibinfo{author}{Avallone, F.}, \bibinfo{author}{Ragni, D.}, \bibinfo{author}{Schram, C.} \& \bibinfo{author}{Sybrand, V. D.~Z.}
\newblock \bibinfo{journal}{\bibinfo{title}{Relevance of quadrupolar sound diffraction on flow-induced noise from porous-coated cylinders}}.
\newblock {\emph{\JournalTitle{Journal of Sound and Vibration}}} \textbf{\bibinfo{volume}{583}}, \bibinfo{pages}{118430} (\bibinfo{year}{2024}).

\bibitem{jones2005benchmark}
\bibinfo{author}{Jones, M.~G.}, \bibinfo{author}{Watson, W.~R.} \& \bibinfo{author}{Parrott, T.~L.}
\newblock \bibinfo{journal}{\bibinfo{title}{Benchmark data for evaluation of aeroacoustic propagation codes with grazing flow}}.
\newblock {\emph{\JournalTitle{AIAA paper}}} \textbf{\bibinfo{volume}{2853}}, \bibinfo{pages}{2005}, \doiprefix\url{10.2514/6.2005-2853} (\bibinfo{year}{2005}).

\bibitem{bonomo}
\bibinfo{author}{Bonomo, L.~A.} \emph{et~al.}
\newblock \emph{\bibinfo{title}{A Comparison of Impedance Eduction Test Rigs with Different Flow Profiles}}.
\newblock \eprint{https://arc.aiaa.org/doi/pdf/10.2514/6.2023-3346}.

\bibitem{auregan2008}
\bibinfo{author}{Aurégan, Y.} \& \bibinfo{author}{Leroux, M.}
\newblock \bibinfo{journal}{\bibinfo{title}{Experimental evidence of an instability over an impedance wall in a duct with flow}}.
\newblock {\emph{\JournalTitle{Journal of Sound and Vibration}}} \textbf{\bibinfo{volume}{317}}, \bibinfo{pages}{432--439}, \doiprefix\url{https://doi.org/10.1016/j.jsv.2008.04.020} (\bibinfo{year}{2008}).

\bibitem{zheng2022}
\bibinfo{author}{Zheng, M.}, \bibinfo{author}{Chen, C.} \& \bibinfo{author}{Li, X.}
\newblock \bibinfo{journal}{\bibinfo{title}{Experimental investigation of factors influencing acoustic liner drag using direct measurement}}.
\newblock {\emph{\JournalTitle{Aerospace Science and Technology}}} \textbf{\bibinfo{volume}{130}}, \bibinfo{pages}{107903}, \doiprefix\url{https://doi.org/10.1016/j.ast.2022.107903} (\bibinfo{year}{2022}).

\bibitem{minotti}
\bibinfo{author}{Minotti, A.}, \bibinfo{author}{Simon, F.} \& \bibinfo{author}{Gantié, F.}
\newblock \bibinfo{journal}{\bibinfo{title}{Characterization of an acoustic liner by means of {L}aser {D}oppler {V}elocimetry in a subsonic flow}}.
\newblock {\emph{\JournalTitle{Aerospace Science and Technology}}} \textbf{\bibinfo{volume}{12}}, \bibinfo{pages}{398--407}, \doiprefix\url{https://doi.org/10.1016/j.ast.2007.09.007} (\bibinfo{year}{2008}).

\bibitem{jing2008}
\bibinfo{author}{Jing, X.}, \bibinfo{author}{Peng, S.} \& \bibinfo{author}{Sun, X.}
\newblock \bibinfo{journal}{\bibinfo{title}{{A straightforward method for wall impedance eduction in a flow duct}}}.
\newblock {\emph{\JournalTitle{The Journal of the Acoustical Society of America}}} \textbf{\bibinfo{volume}{124}}, \doiprefix\url{10.1121/1.2932256} (\bibinfo{year}{2008}).

\bibitem{eversman2011}
\bibinfo{author}{Eversman, W.} \& \bibinfo{author}{Gallman, J.~M.}
\newblock \bibinfo{journal}{\bibinfo{title}{Impedance eduction with an extended search procedure}}.
\newblock {\emph{\JournalTitle{AIAA journal}}} \textbf{\bibinfo{volume}{49}}, \bibinfo{pages}{1960--1970}, \doiprefix\url{10.2514/1.J050831} (\bibinfo{year}{2011}).

\bibitem{primus2013adjoint}
\bibinfo{author}{Primus, J.}, \bibinfo{author}{Piot, E.} \& \bibinfo{author}{Simon, F.}
\newblock \bibinfo{journal}{\bibinfo{title}{An adjoint-based method for liner impedance eduction: Validation and numerical investigation}}.
\newblock {\emph{\JournalTitle{Journal of Sound and Vibration}}} \textbf{\bibinfo{volume}{332}}, \bibinfo{pages}{58--75}, \doiprefix\url{10.1016/j.jsv.2012.07.051} (\bibinfo{year}{2013}).

\bibitem{Ingard1959}
\bibinfo{author}{Ingard, U.}
\newblock \bibinfo{journal}{\bibinfo{title}{{Influence of Fluid Motion Past a Plane Boundary on Sound Reflection, Absorption, and Transmission}}}.
\newblock {\emph{\JournalTitle{The Journal of the Acoustical Society of America}}} \textbf{\bibinfo{volume}{31}}, \bibinfo{pages}{1035--1036}, \doiprefix\url{10.1121/1.1907805} (\bibinfo{year}{1959}).

\bibitem{Myers1980OnFlow}
\bibinfo{author}{Myers, M.~K.}
\newblock \bibinfo{journal}{\bibinfo{title}{{On the acoustic boundary condition in the presence of flow}}}.
\newblock {\emph{\JournalTitle{Journal of Sound and Vibration}}} \textbf{\bibinfo{volume}{71}}, \doiprefix\url{10.1016/0022-460X(80)90424-1} (\bibinfo{year}{1980}).

\bibitem{jones2010effects}
\bibinfo{author}{Jones, M.}, \bibinfo{author}{Watson, W.} \& \bibinfo{author}{Nark, D.}
\newblock \bibinfo{title}{Effects of flow profile on educed acoustic liner impedance}.
\newblock In \emph{\bibinfo{booktitle}{16th AIAA/CEAS Aeroacoustics Conference}}, \bibinfo{pages}{3763}, \doiprefix\url{10.2514/6.2010-3763} (\bibinfo{year}{2010}).

\bibitem{Quintino2025}
\bibinfo{author}{Quintino, N.~T.} \emph{et~al.}
\newblock \bibinfo{journal}{\bibinfo{title}{Comparison of impedance eduction test rigs with different boundary-layer profiles}}.
\newblock {\emph{\JournalTitle{AIAA Journal}}} \doiprefix\url{10.2514/1.J065173} (\bibinfo{year}{2025}).

\bibitem{brown}
\bibinfo{author}{Brown, M.}, \bibinfo{author}{Jones, M.} \& \bibinfo{author}{Watson, W.}
\newblock \bibinfo{title}{Uncertainty analysis of the grazing flow impedance tube}.
\newblock In \emph{\bibinfo{booktitle}{18th AIAA/CEAS Aeroacoustics Conference (33rd AIAA Aeroacoustics Conference)}}, \bibinfo{pages}{1--12}, \doiprefix\url{10.2514/6.2012-2296} (\bibinfo{publisher}{American Institute of Aeronautics and Astronautics (AIAA)}, \bibinfo{year}{2012}).
\newblock \eprint{https://arc.aiaa.org/doi/pdf/10.2514/6.2012-2296}.

\bibitem{rroncen2018}
\bibinfo{author}{Roncen, R.}, \bibinfo{author}{Méry, F.}, \bibinfo{author}{Piot, E.} \& \bibinfo{author}{Simon, F.}
\newblock \bibinfo{journal}{\bibinfo{title}{Statistical inference method for liner impedance eduction with a shear grazing flow}}.
\newblock {\emph{\JournalTitle{AIAA Journal}}} \textbf{\bibinfo{volume}{57}}, \bibinfo{pages}{1055--1065}, \doiprefix\url{10.2514/1.J057559} (\bibinfo{year}{2019}).
\newblock \eprint{https://doi.org/10.2514/1.J057559}.

\bibitem{zhou}
\bibinfo{author}{Zhou, L.} \& \bibinfo{author}{Bodén, H.}
\newblock \bibinfo{journal}{\bibinfo{title}{A systematic uncertainty analysis for liner impedance eduction technology}}.
\newblock {\emph{\JournalTitle{Journal of Sound and Vibration}}} \textbf{\bibinfo{volume}{356}}, \bibinfo{pages}{86--99}, \doiprefix\url{https://doi.org/10.1016/j.jsv.2015.07.001} (\bibinfo{year}{2015}).

\bibitem{howerton2017}
\bibinfo{author}{Howerton, B.~M.} \& \bibinfo{author}{G., M.}
\newblock \bibinfo{journal}{\bibinfo{title}{A conventional liner acoustic drag interaction benchmark database}}.
\newblock {\emph{\JournalTitle{23rd AIAA/CEAS Aeroacoustics Conference}}} \doiprefix\url{10.2514/6.2017-4190}.

\bibitem{jasinski}
\bibinfo{author}{Jasinski, C.} \& \bibinfo{author}{Corke, T.}
\newblock \bibinfo{journal}{\bibinfo{title}{Mechanism for increased viscous drag over porous sheet acoustic liners}}.
\newblock {\emph{\JournalTitle{AIAA Journal}}} \textbf{\bibinfo{volume}{58}}, \bibinfo{pages}{3393--3404}, \doiprefix\url{10.2514/1.J059039} (\bibinfo{year}{2020}).
\newblock \eprint{https://doi.org/10.2514/1.J059039}.

\bibitem{leon2019}
\bibinfo{author}{L{\'e}on, O.}, \bibinfo{author}{M{\'e}ry, F.}, \bibinfo{author}{Piot, E.} \& \bibinfo{author}{Conte, C.}
\newblock \bibinfo{journal}{\bibinfo{title}{{Near-wall aerodynamic response of an acoustic liner to harmonic excitation with grazing flow}}}.
\newblock {\emph{\JournalTitle{Experiments in Fluid}}} \textbf{\bibinfo{volume}{60}}, \doiprefix\url{10.1007/s00348-019-2791-5} (\bibinfo{year}{2019}).

\bibitem{lafont2021}
\bibinfo{author}{Lafont, V.}, \bibinfo{author}{M{\'e}ry, F.}, \bibinfo{author}{Reulet, P.} \& \bibinfo{author}{Simon, F.}
\newblock \bibinfo{journal}{\bibinfo{title}{{Surface temperature measurement of acoustic liners in presence of grazing flow and thermal gradient}}}.
\newblock {\emph{\JournalTitle{Experiments in Fluid}}} \textbf{\bibinfo{volume}{62}}, \doiprefix\url{10.1007/s00348-021-03184-w} (\bibinfo{year}{2021}).

\bibitem{zhen2025resonance}
\bibinfo{author}{Zhen, N.}, \bibinfo{author}{Huang, R.-R.}, \bibinfo{author}{Fan, S.-W.}, \bibinfo{author}{Wang, Y.-F.} \& \bibinfo{author}{Wang, Y.-S.}
\newblock \bibinfo{journal}{\bibinfo{title}{Resonance-based acoustic ventilated metamaterials for sound insulation}}.
\newblock {\emph{\JournalTitle{npj Acoustics}}} \textbf{\bibinfo{volume}{1}}, \bibinfo{pages}{7} (\bibinfo{year}{2025}).

\bibitem{kumar2019present}
\bibinfo{author}{Kumar, S.} \& \bibinfo{author}{Lee, H.~P.}
\newblock \bibinfo{title}{The present and future role of acoustic metamaterials for architectural and urban noise mitigations}.
\newblock In \emph{\bibinfo{booktitle}{Acoustics}}, vol.~\bibinfo{volume}{1}, \bibinfo{pages}{590--607} (\bibinfo{year}{2019}).

\bibitem{ingard2009noise}
\bibinfo{author}{Ingard, U.}
\newblock \emph{\bibinfo{title}{Noise reduction analysis}} (\bibinfo{publisher}{Jones \& Bartlett Publishers}, \bibinfo{year}{2009}).

\bibitem{stewart1926tube}
\bibinfo{author}{Stewart, G.~W.}
\newblock \bibinfo{journal}{\bibinfo{title}{The tube as a branch of an acoustic conduct: The special case of the {Q}uincke tube}}.
\newblock {\emph{\JournalTitle{Phys. Rev.}}} \textbf{\bibinfo{volume}{27}}, \bibinfo{pages}{494--498}, \doiprefix\url{10.1103/PhysRev.27.494} (\bibinfo{year}{1926}).

\bibitem{poirier2011use}
\bibinfo{author}{Poirier, B.}, \bibinfo{author}{Maury, C.} \& \bibinfo{author}{Ville, J.-M.}
\newblock \bibinfo{journal}{\bibinfo{title}{The use of {H}erschel--{Q}uincke tubes to improve the efficiency of lined ducts}}.
\newblock {\emph{\JournalTitle{Applied acoustics}}} \textbf{\bibinfo{volume}{72}}, \bibinfo{pages}{78--88} (\bibinfo{year}{2011}).

\bibitem{papathanasiou2022herschel}
\bibinfo{author}{Papathanasiou, T.}, \bibinfo{author}{Tsolakis, E.}, \bibinfo{author}{Spitas, V.} \& \bibinfo{author}{Movchan, A.}
\newblock \bibinfo{journal}{\bibinfo{title}{The {H}erschel--{Q}uincke tube with modulated branches}}.
\newblock {\emph{\JournalTitle{Philosophical Transactions of the Royal Society A}}} \textbf{\bibinfo{volume}{380}}, \bibinfo{pages}{20220074} (\bibinfo{year}{2022}).

\bibitem{ko1971sound}
\bibinfo{author}{Ko, S.-H.}
\newblock \bibinfo{journal}{\bibinfo{title}{Sound attenuation in lined rectangular ducts with flow and its application to the reduction of aircraft engine noise}}.
\newblock {\emph{\JournalTitle{The Journal of the Acoustical Society of America}}} \textbf{\bibinfo{volume}{50}}, \bibinfo{pages}{1418--1432} (\bibinfo{year}{1971}).

\bibitem{arenas2010recent}
\bibinfo{author}{Arenas, J.~P.} \& \bibinfo{author}{Crocker, M.~J.}
\newblock \bibinfo{journal}{\bibinfo{title}{Recent trends in porous sound-absorbing materials}}.
\newblock {\emph{\JournalTitle{Sound \& vibration}}} \textbf{\bibinfo{volume}{44}}, \bibinfo{pages}{12--18} (\bibinfo{year}{2010}).

\bibitem{liang2012extreme}
\bibinfo{author}{Liang, Z.} \& \bibinfo{author}{Li, J.}
\newblock \bibinfo{journal}{\bibinfo{title}{Extreme acoustic metamaterial by coiling up space}}.
\newblock {\emph{\JournalTitle{Physical Review Letters}}} \textbf{\bibinfo{volume}{108}}, \bibinfo{pages}{114301} (\bibinfo{year}{2012}).

\bibitem{ghaffarivardavagh2019ultra}
\bibinfo{author}{Ghaffarivardavagh, R.}, \bibinfo{author}{Nikolajczyk, J.}, \bibinfo{author}{Anderson, S.} \& \bibinfo{author}{Zhang, X.}
\newblock \bibinfo{journal}{\bibinfo{title}{Ultra-open acoustic metamaterial silencer based on fano-like interference}}.
\newblock {\emph{\JournalTitle{Physical Review B}}} \textbf{\bibinfo{volume}{99}}, \bibinfo{pages}{024302} (\bibinfo{year}{2019}).

\bibitem{sun2020broadband}
\bibinfo{author}{Sun, M.}, \bibinfo{author}{Fang, X.}, \bibinfo{author}{Mao, D.}, \bibinfo{author}{Wang, X.} \& \bibinfo{author}{Li, Y.}
\newblock \bibinfo{journal}{\bibinfo{title}{Broadband acoustic ventilation barriers}}.
\newblock {\emph{\JournalTitle{Physical Review Applied}}} \textbf{\bibinfo{volume}{13}}, \bibinfo{pages}{044028} (\bibinfo{year}{2020}).

\bibitem{tang2022broadband}
\bibinfo{author}{Tang, Y.}, \bibinfo{author}{Liang, B.} \& \bibinfo{author}{Lin, S.}
\newblock \bibinfo{journal}{\bibinfo{title}{Broadband ventilated meta-barrier based on the synergy of mode superposition and consecutive fano resonances}}.
\newblock {\emph{\JournalTitle{The Journal of the Acoustical Society of America}}} \textbf{\bibinfo{volume}{152}}, \bibinfo{pages}{2412--2418} (\bibinfo{year}{2022}).

\bibitem{zhu2023nonlocal}
\bibinfo{author}{Zhu, Y.}, \bibinfo{author}{Dong, R.}, \bibinfo{author}{Mao, D.}, \bibinfo{author}{Wang, X.} \& \bibinfo{author}{Li, Y.}
\newblock \bibinfo{journal}{\bibinfo{title}{Nonlocal ventilating metasurfaces}}.
\newblock {\emph{\JournalTitle{Physical Review Applied}}} \textbf{\bibinfo{volume}{19}}, \bibinfo{pages}{014067} (\bibinfo{year}{2023}).

\bibitem{liu2017spiral}
\bibinfo{author}{Liu, C.}, \bibinfo{author}{Xia, B.} \& \bibinfo{author}{Yu, D.}
\newblock \bibinfo{journal}{\bibinfo{title}{The spiral-labyrinthine acoustic metamaterial by coiling up space}}.
\newblock {\emph{\JournalTitle{Physics Letters A}}} \textbf{\bibinfo{volume}{381}}, \bibinfo{pages}{3112--3118} (\bibinfo{year}{2017}).

\bibitem{xiao2021ventilated}
\bibinfo{author}{Xiao, Z.}, \bibinfo{author}{Gao, P.}, \bibinfo{author}{Wang, D.}, \bibinfo{author}{He, X.} \& \bibinfo{author}{Wu, L.}
\newblock \bibinfo{journal}{\bibinfo{title}{Ventilated metamaterials for broadband sound insulation and tunable transmission at low frequency}}.
\newblock {\emph{\JournalTitle{Extreme Mechanics Letters}}} \textbf{\bibinfo{volume}{46}}, \bibinfo{pages}{101348} (\bibinfo{year}{2021}).

\bibitem{chen2022broadband}
\bibinfo{author}{Chen, A.}, \bibinfo{author}{Zhao, X.}, \bibinfo{author}{Yang, Z.}, \bibinfo{author}{Anderson, S.} \& \bibinfo{author}{Zhang, X.}
\newblock \bibinfo{journal}{\bibinfo{title}{Broadband labyrinthine acoustic insulator}}.
\newblock {\emph{\JournalTitle{Physical Review Applied}}} \textbf{\bibinfo{volume}{18}}, \bibinfo{pages}{064057} (\bibinfo{year}{2022}).

\bibitem{lee2025compact}
\bibinfo{author}{Lee, I.}, \bibinfo{author}{Han, I.} \& \bibinfo{author}{Yoon, G.}
\newblock \bibinfo{journal}{\bibinfo{title}{Compact acoustic metamaterials based on azimuthal labyrinthine channels for broadband low-frequency soundproofing and ventilation}}.
\newblock {\emph{\JournalTitle{Applied Acoustics}}} \textbf{\bibinfo{volume}{228}}, \bibinfo{pages}{110273} (\bibinfo{year}{2025}).

\bibitem{liu2021three}
\bibinfo{author}{Liu, C.} \emph{et~al.}
\newblock \bibinfo{journal}{\bibinfo{title}{Three-dimensional soundproof acoustic metacage}}.
\newblock {\emph{\JournalTitle{Physical Review Letters}}} \textbf{\bibinfo{volume}{127}}, \bibinfo{pages}{084301} (\bibinfo{year}{2021}).

\bibitem{krasikova2023metahouse}
\bibinfo{author}{Krasikova, M.} \emph{et~al.}
\newblock \bibinfo{journal}{\bibinfo{title}{Metahouse: noise-insulating chamber based on periodic structures}}.
\newblock {\emph{\JournalTitle{Advanced Materials Technologies}}} \textbf{\bibinfo{volume}{8}}, \bibinfo{pages}{2200711} (\bibinfo{year}{2023}).

\bibitem{meng2023subwavelength}
\bibinfo{author}{Meng, Y.} \emph{et~al.}
\newblock \bibinfo{journal}{\bibinfo{title}{Subwavelength broadband perfect absorption for unidimensional open-duct problems}}.
\newblock {\emph{\JournalTitle{Advanced Materials Technologies}}} \textbf{\bibinfo{volume}{8}}, \bibinfo{pages}{2201909} (\bibinfo{year}{2023}).

\bibitem{yang2025phase}
\bibinfo{author}{Yang, Z.}, \bibinfo{author}{Chen, A.}, \bibinfo{author}{Xie, X.}, \bibinfo{author}{Anderson, S.~W.} \& \bibinfo{author}{Zhang, X.}
\newblock \bibinfo{journal}{\bibinfo{title}{Phase gradient ultra open metamaterials for broadband acoustic silencing}}.
\newblock {\emph{\JournalTitle{Scientific Reports}}} \textbf{\bibinfo{volume}{15}}, \bibinfo{pages}{21434} (\bibinfo{year}{2025}).

\bibitem{yang2017optimal}
\bibinfo{author}{Yang, M.}, \bibinfo{author}{Chen, S.}, \bibinfo{author}{Fu, C.} \& \bibinfo{author}{Sheng, P.}
\newblock \bibinfo{journal}{\bibinfo{title}{Optimal sound-absorbing structures}}.
\newblock {\emph{\JournalTitle{Materials Horizons}}} \textbf{\bibinfo{volume}{4}}, \bibinfo{pages}{673--680} (\bibinfo{year}{2017}).

\bibitem{li2022frequency}
\bibinfo{author}{Li, X.}, \bibinfo{author}{Zhang, H.}, \bibinfo{author}{Tian, H.}, \bibinfo{author}{Huang, Y.} \& \bibinfo{author}{Wang, L.}
\newblock \bibinfo{journal}{\bibinfo{title}{Frequency-tunable sound insulation via a reconfigurable and ventilated acoustic metamaterial}}.
\newblock {\emph{\JournalTitle{Journal of Physics D: Applied Physics}}} \textbf{\bibinfo{volume}{55}}, \bibinfo{pages}{495108} (\bibinfo{year}{2022}).

\bibitem{wen2023origami}
\bibinfo{author}{Wen, G.} \emph{et~al.}
\newblock \bibinfo{journal}{\bibinfo{title}{Origami-based acoustic metamaterial for tunable and broadband sound attenuation}}.
\newblock {\emph{\JournalTitle{International Journal of Mechanical Sciences}}} \textbf{\bibinfo{volume}{239}}, \bibinfo{pages}{107872} (\bibinfo{year}{2023}).

\bibitem{zhang-et-al:jnm2020}
\bibinfo{author}{Shuaizhong~Zhang, P.~O., Ye~Wang} \& \bibinfo{author}{den Toonder, J.}
\newblock \bibinfo{journal}{\bibinfo{title}{{A concise review of microfluidic particle manipulation methods}}}.
\newblock {\emph{\JournalTitle{Microfluidics and Nanofluidics}}} \textbf{\bibinfo{volume}{24}}, \bibinfo{pages}{20}, \doiprefix\url{https://doi.org/10.1007/s10404-020-2328-5} (\bibinfo{year}{2020}).

\bibitem{godary-et-al:acs2025}
\bibinfo{author}{Godary, T.} \emph{et~al.}
\newblock \bibinfo{journal}{\bibinfo{title}{{Acoustofluidics: Technology Advances and Applications from 2022 to 2024}}}.
\newblock {\emph{\JournalTitle{Analytical Chemistry}}} \textbf{\bibinfo{volume}{97}}, \bibinfo{pages}{24}, \doiprefix\url{https://doi.org/10.1021/acs.analchem.4c06803} (\bibinfo{year}{2025}).

\bibitem{ding-et-al:loc2016}
\bibinfo{author}{Ding, X.} \emph{et~al.}
\newblock \bibinfo{journal}{\bibinfo{title}{{Surface acoustic wave microfluidics}}}.
\newblock {\emph{\JournalTitle{Lab on a Chip}}} \textbf{\bibinfo{volume}{13}}, \bibinfo{pages}{3626–3649}, \doiprefix\url{DOI: 10.1039/c3lc50361e} (\bibinfo{year}{2016}).

\bibitem{guo-et-al:pnas2016}
\bibinfo{author}{Guo, F.} \emph{et~al.}
\newblock \bibinfo{journal}{\bibinfo{title}{{Three-dimensional manipulation of single cells using surface acoustic waves}}}.
\newblock {\emph{\JournalTitle{Proc. Natl. Acad. Sci.}}} \textbf{\bibinfo{volume}{113}}, \bibinfo{pages}{1522--1527}, \doiprefix\url{https://doi.org/10.1073/pnas.1524813113 (2016)} (\bibinfo{year}{2016}).

\bibitem{wu-et-al:micronano2019}
\bibinfo{author}{Wu, M.} \emph{et~al.}
\newblock \bibinfo{journal}{\bibinfo{title}{{Acoustofluidic separation of cells and particles}}}.
\newblock {\emph{\JournalTitle{Microsystems \& Nanoengineering}}} \textbf{\bibinfo{volume}{5}}, \bibinfo{pages}{18}, \doiprefix\url{https://doi.org/10.1038/s41378-019-0064-3} (\bibinfo{year}{2019}).

\bibitem{shen-et-al:sciadv2024}
\bibinfo{author}{Shen, L.} \emph{et~al.}
\newblock \bibinfo{journal}{\bibinfo{title}{{Acousto-dielectric tweezers enable independent manipulation of multiple particles}}}.
\newblock {\emph{\JournalTitle{Science Advances}}} \textbf{\bibinfo{volume}{10}}, \bibinfo{pages}{10}, \doiprefix\url{DOI: 10.1126/sciadv.ado8992} (\bibinfo{year}{2024}).

\bibitem{bruus:loac2013}
\bibinfo{author}{Bruus, H.}
\newblock \bibinfo{journal}{\bibinfo{title}{{Acoustofluidics 7: The acoustic radiation force on small particles}}}.
\newblock {\emph{\JournalTitle{Lab on a Chip}}} \textbf{\bibinfo{volume}{12}}, \bibinfo{pages}{1014–1021}, \doiprefix\url{DOI https://doi.org/10.1039/C2LC21068A} (\bibinfo{year}{2013}).

\bibitem{ellen:loac2018}
\bibinfo{author}{Cesewski, E.} \emph{et~al.}
\newblock \bibinfo{journal}{\bibinfo{title}{{Additive manufacturing of three-dimensional (3{D}) microfluidic-based microelectromechanical systems (MEMS) for acoustofluidic applications}}}.
\newblock {\emph{\JournalTitle{Lab on a Chip}}} \textbf{\bibinfo{volume}{18}}, \bibinfo{pages}{2087–2098}, \doiprefix\url{DOI https://doi.org/10.1039/C8LC00427G} (\bibinfo{year}{2018}).

\bibitem{fei:prappl2020}
\bibinfo{author}{Li, F.} \emph{et~al.}
\newblock \bibinfo{journal}{\bibinfo{title}{{Phononic-Crystal-Enabled Dynamic Manipulation of Microparticles and Cells in an Acoustofluidic Channel}}}.
\newblock {\emph{\JournalTitle{Phys. Rev. Applied}}} \textbf{\bibinfo{volume}{13}}, \bibinfo{pages}{044077}, \doiprefix\url{10.1103/PhysRevApplied.13.044077} (\bibinfo{year}{2020}).

\bibitem{huang:apl2022}
\bibinfo{author}{Huang, L.} \emph{et~al.}
\newblock \bibinfo{journal}{\bibinfo{title}{{Phononic crystal-induced standing Lamb wave for the translation of subwavelength microparticles}}}.
\newblock {\emph{\JournalTitle{Appl. Phys. Lett.}}} \textbf{\bibinfo{volume}{121}}, \bibinfo{pages}{023505}, \doiprefix\url{https://doi.org/10.1063/5.0098468} (\bibinfo{year}{2022}).

\bibitem{zeighami:philtransA2024}
\bibinfo{author}{Zeighami, F.} \emph{et~al.}
\newblock \bibinfo{journal}{\bibinfo{title}{{Elastic metasurfaces for Scholte–Stoneley wave control}}}.
\newblock {\emph{\JournalTitle{Philosophical Transactions of the Royal Society A: Mathematical, Physical and Engineering Sciences}}} \textbf{\bibinfo{volume}{382}}, \bibinfo{pages}{20230365}, \doiprefix\url{https://doi.org/10.1098/rsta.2023.0365} (\bibinfo{year}{2024}).

\bibitem{surappa:natcommun2025}
\bibinfo{author}{Surappa, S.} \emph{et~al.}
\newblock \bibinfo{journal}{\bibinfo{title}{{Dynamically reconfigurable acoustofluidic metasurface for subwavelength particle manipulation and assembly}}}.
\newblock {\emph{\JournalTitle{Nature Communications}}} \textbf{\bibinfo{volume}{16}}, \bibinfo{pages}{1--10}, \doiprefix\url{https://doi.org/10.1038/s41467-024-55337-0} (\bibinfo{year}{2025}).

\bibitem{kadic20193d}
\bibinfo{author}{Kadic, M.}, \bibinfo{author}{Milton, G.~W.}, \bibinfo{author}{van Hecke, M.} \& \bibinfo{author}{Wegener, M.}
\newblock \bibinfo{journal}{\bibinfo{title}{3{D} metamaterials}}.
\newblock {\emph{\JournalTitle{Nature Reviews Physics}}} \textbf{\bibinfo{volume}{1}}, \bibinfo{pages}{198--210} (\bibinfo{year}{2019}).

\bibitem{jin20252024}
\bibinfo{author}{Jin, Y.} \emph{et~al.}
\newblock \bibinfo{journal}{\bibinfo{title}{The 2024 phononic crystals roadmap}}.
\newblock {\emph{\JournalTitle{Journal of Physics D: Applied Physics}}} \textbf{\bibinfo{volume}{58}}, \bibinfo{pages}{113001} (\bibinfo{year}{2025}).

\bibitem{chen2020mapping}
\bibinfo{author}{Chen, Y.}, \bibinfo{author}{Frenzel, T.}, \bibinfo{author}{Guenneau, S.}, \bibinfo{author}{Kadic, M.} \& \bibinfo{author}{Wegener, M.}
\newblock \bibinfo{journal}{\bibinfo{title}{Mapping acoustical activity in 3{D} chiral mechanical metamaterials onto micropolar continuum elasticity}}.
\newblock {\emph{\JournalTitle{Journal of the Mechanics and Physics of Solids}}} \textbf{\bibinfo{volume}{137}}, \bibinfo{pages}{103877} (\bibinfo{year}{2020}).

\bibitem{chen2021roton}
\bibinfo{author}{Chen, Y.}, \bibinfo{author}{Kadic, M.} \& \bibinfo{author}{Wegener, M.}
\newblock \bibinfo{journal}{\bibinfo{title}{Roton-like acoustical dispersion relations in 3{D} metamaterials}}.
\newblock {\emph{\JournalTitle{Nature Communications}}} \textbf{\bibinfo{volume}{12}}, \bibinfo{pages}{3278} (\bibinfo{year}{2021}).

\bibitem{craster2023mechanical}
\bibinfo{author}{Craster, R.}, \bibinfo{author}{Guenneau, S.}, \bibinfo{author}{Kadic, M.} \& \bibinfo{author}{Wegener, M.}
\newblock \bibinfo{journal}{\bibinfo{title}{Mechanical metamaterials}}.
\newblock {\emph{\JournalTitle{Reports on Progress in Physics}}} \textbf{\bibinfo{volume}{86}}, \bibinfo{pages}{094501} (\bibinfo{year}{2023}).

\bibitem{chen2025nonlocal}
\bibinfo{author}{Chen, Y.}, \bibinfo{author}{Fleury, R.}, \bibinfo{author}{Seppecher, P.}, \bibinfo{author}{Hu, G.} \& \bibinfo{author}{Wegener, M.}
\newblock \bibinfo{journal}{\bibinfo{title}{Nonlocal metamaterials and metasurfaces}}.
\newblock {\emph{\JournalTitle{Nature Reviews Physics}}} \textbf{\bibinfo{volume}{7}}, \bibinfo{pages}{299--312} (\bibinfo{year}{2025}).

\bibitem{nassar2020nonreciprocity}
\bibinfo{author}{Nassar, H.} \emph{et~al.}
\newblock \bibinfo{journal}{\bibinfo{title}{Nonreciprocity in acoustic and elastic materials}}.
\newblock {\emph{\JournalTitle{Nat. Rev. Mater.}}} \textbf{\bibinfo{volume}{5}}, \bibinfo{pages}{667--685} (\bibinfo{year}{2020}).

\bibitem{chen2020high}
\bibinfo{author}{Chen, Y.} \emph{et~al.}
\newblock \bibinfo{journal}{\bibinfo{title}{High-frequency gravitational-wave detection using a chiral resonant mechanical element and a short unstable optical cavity}}.
\newblock {\emph{\JournalTitle{arXiv preprint arXiv:2007.07974}}}  (\bibinfo{year}{2020}).

\bibitem{cummer2016controlling}
\bibinfo{author}{Cummer, S.~A.}, \bibinfo{author}{Christensen, J.} \& \bibinfo{author}{Al{\`u}, A.}
\newblock \bibinfo{journal}{\bibinfo{title}{Controlling sound with acoustic metamaterials}}.
\newblock {\emph{\JournalTitle{Nature Reviews Materials}}} \textbf{\bibinfo{volume}{1}}, \bibinfo{pages}{1--13} (\bibinfo{year}{2016}).

\bibitem{zhang2021wave}
\bibinfo{author}{Zhang, S.-Y.}, \bibinfo{author}{Yan, D.-J.}, \bibinfo{author}{Wang, Y.-S.}, \bibinfo{author}{Wang, Y.-F.} \& \bibinfo{author}{Laude, V.}
\newblock \bibinfo{journal}{\bibinfo{title}{Wave propagation in one-dimensional fluid-saturated porous phononic crystals with partial-open pore interfaces}}.
\newblock {\emph{\JournalTitle{International Journal of Mechanical Sciences}}} \textbf{\bibinfo{volume}{195}}, \bibinfo{pages}{106227} (\bibinfo{year}{2021}).

\bibitem{fleury:natcommun2016}
\bibinfo{author}{Fleury, R.}, \bibinfo{author}{Khanikaev, A.~B.} \& \bibinfo{author}{Alù, A.}
\newblock \bibinfo{journal}{\bibinfo{title}{{Floquet topological insulators for sound}}}.
\newblock {\emph{\JournalTitle{Nature Communications}}} \textbf{\bibinfo{volume}{7}}, \bibinfo{pages}{11744}, \doiprefix\url{https://doi.org/10.1038/ncomms11744} (\bibinfo{year}{2016}).

\bibitem{huber:natphys2016}
\bibinfo{author}{Huber, S.~D.}
\newblock \bibinfo{journal}{\bibinfo{title}{{Topological mechanics}}}.
\newblock {\emph{\JournalTitle{Nature Physics}}} \textbf{\bibinfo{volume}{12}}, \bibinfo{pages}{621–623}, \doiprefix\url{https://doi.org/10.1038/nphys3801} (\bibinfo{year}{2016}).

\bibitem{khanikaev:natmat2013}
\bibinfo{author}{Khanikaev, A.~B.} \emph{et~al.}
\newblock \bibinfo{journal}{\bibinfo{title}{{Photonic topological insulators}}}.
\newblock {\emph{\JournalTitle{Nature Materials}}} \textbf{\bibinfo{volume}{12}}, \bibinfo{pages}{233–239}, \doiprefix\url{https://doi.org/10.1038/nmat3520} (\bibinfo{year}{2013}).

\bibitem{BertoldiPRL2015}
\bibinfo{author}{Wang, P.}, \bibinfo{author}{Lu, L.} \& \bibinfo{author}{Bertoldi, K.}
\newblock \bibinfo{journal}{\bibinfo{title}{{Topological Phononic Crystals with One-Way Elastic Edge Waves}}}.
\newblock {\emph{\JournalTitle{Physical Review Letters}}} \textbf{\bibinfo{volume}{115}}, \bibinfo{pages}{104302}, \doiprefix\url{https://doi.org/10.1103/PhysRevLett.115.104302} (\bibinfo{year}{2015}).

\bibitem{MaNatRevPhys2019}
\bibinfo{author}{Ma, G.}, \bibinfo{author}{Xiao, M.} \& \bibinfo{author}{Chan, C.~T.}
\newblock \bibinfo{journal}{\bibinfo{title}{{Topological phases in acoustic and mechanical systems}}}.
\newblock {\emph{\JournalTitle{Nature Reviews Physics}}} \textbf{\bibinfo{volume}{1}}, \bibinfo{pages}{281--294}, \doiprefix\url{https://doi.org/10.1038/nmat3520} (\bibinfo{year}{2019}).

\bibitem{berry:philtrans1984}
\bibinfo{author}{Berry, M.~V.}
\newblock \bibinfo{journal}{\bibinfo{title}{{Quantal phase factors accompanying adiabatic changes}}}.
\newblock {\emph{\JournalTitle{Proceedings of the Royal Society of London. A. Mathematical and Physical Sciences}}} \textbf{\bibinfo{volume}{392}}, \doiprefix\url{https://doi.org/10.1098/rspa.1984.0023} (\bibinfo{year}{1984}).

\bibitem{zak:prl1989}
\bibinfo{author}{Zak, J.}
\newblock \bibinfo{journal}{\bibinfo{title}{{Berry’s phase for energy bands in solids}}}.
\newblock {\emph{\JournalTitle{Phys. Rev. Lett.}}} \textbf{\bibinfo{volume}{62}}, \bibinfo{pages}{2747}, \doiprefix\url{https://doi.org/10.1103/PhysRevLett.62.2747} (\bibinfo{year}{1989}).

\bibitem{jiang:prb2019}
\bibinfo{author}{Jiang, W.} \emph{et~al.}
\newblock \bibinfo{journal}{\bibinfo{title}{{Topological band evolution between Lieb and kagome lattices}}}.
\newblock {\emph{\JournalTitle{Phys. Rev. B}}} \textbf{\bibinfo{volume}{99}}, \bibinfo{pages}{125131}, \doiprefix\url{https://doi.org/10.1103/PhysRevB.99.125131} (\bibinfo{year}{2019}).

\bibitem{martínez:prb2022}
\bibinfo{author}{Martínez, J. A.~I.}, \bibinfo{author}{Laforge, N.}, \bibinfo{author}{Kadic, M.} \& \bibinfo{author}{Laude, V.}
\newblock \bibinfo{journal}{\bibinfo{title}{{Topological waves guided by a glide-reflection symmetric crystal interface}}}.
\newblock {\emph{\JournalTitle{Phys. Rev. B}}} \textbf{\bibinfo{volume}{106}}, \bibinfo{pages}{0643041}, \doiprefix\url{DOI: 10.1103/PhysRevB.106.064304} (\bibinfo{year}{2022}).

\bibitem{deponti:natcommun2024}
\bibinfo{author}{Ponti, J. M.~D.} \emph{et~al.}
\newblock \bibinfo{journal}{\bibinfo{title}{{Localized topological states beyond Fano resonances via counter-propagating wave mode conversion in piezoelectric microelectromechanical devices}}}.
\newblock {\emph{\JournalTitle{Nature Communications}}} \textbf{\bibinfo{volume}{15}}, \bibinfo{pages}{9617}, \doiprefix\url{https://doi.org/10.1038/s41467-024-53925-8} (\bibinfo{year}{2024}).

\bibitem{XueNatMat2019}
\bibinfo{author}{Xue, H.}, \bibinfo{author}{Yang, Y.}, \bibinfo{author}{Gao, F.}, \bibinfo{author}{Chong, Y.} \& \bibinfo{author}{Zhang, B.}
\newblock \bibinfo{journal}{\bibinfo{title}{{Acoustic higher-order topological insulator on a kagome lattice}}}.
\newblock {\emph{\JournalTitle{Nature Materials}}} \textbf{\bibinfo{volume}{18}}, \bibinfo{pages}{108--112}, \doiprefix\url{https://doi.org/10.1038/s41563-018-0251-x} (\bibinfo{year}{2019}).

\bibitem{NiNatMAt2019}
\bibinfo{author}{Ni, X.}, \bibinfo{author}{Weiner, M.}, \bibinfo{author}{Alù, A.} \& \bibinfo{author}{Khanikaev, A.~B.}
\newblock \bibinfo{journal}{\bibinfo{title}{{Observation of higher-order topological acoustic states protected by generalized chiral symmetry}}}.
\newblock {\emph{\JournalTitle{Nature Materials}}} \textbf{\bibinfo{volume}{18}}, \bibinfo{pages}{113--120}, \doiprefix\url{https://doi.org/10.1038/s41563-018-0252-9} (\bibinfo{year}{2019}).

\bibitem{LuoNatMat2021}
\bibinfo{author}{Luo, L.} \emph{et~al.}
\newblock \bibinfo{journal}{\bibinfo{title}{{Observation of a phononic higher-order Weyl semimetal}}}.
\newblock {\emph{\JournalTitle{Nature Materials}}} \textbf{\bibinfo{volume}{20}}, \bibinfo{pages}{794–799}, \doiprefix\url{https://doi.org/10.1038/s41563-021-00985-6} (\bibinfo{year}{2021}).

\bibitem{MaPRL2024}
\bibinfo{author}{Ma, Q.} \emph{et~al.}
\newblock \bibinfo{journal}{\bibinfo{title}{{Observation of Higher-Order Nodal-Line Semimetal in Phononic Crystals}}}.
\newblock {\emph{\JournalTitle{Physical Review Letters}}} \textbf{\bibinfo{volume}{132}}, \bibinfo{pages}{066601}, \doiprefix\url{https://doi.org/10.1103/PhysRevLett.132.066601} (\bibinfo{year}{2024}).

\bibitem{yang2015topological}
\bibinfo{author}{Yang, Z.} \emph{et~al.}
\newblock \bibinfo{journal}{\bibinfo{title}{Topological acoustics}}.
\newblock {\emph{\JournalTitle{Physical review letters}}} \textbf{\bibinfo{volume}{114}}, \bibinfo{pages}{114301} (\bibinfo{year}{2015}).

\bibitem{lu2016valley}
\bibinfo{author}{Lu, J.}, \bibinfo{author}{Qiu, C.}, \bibinfo{author}{Ke, M.} \& \bibinfo{author}{Liu, Z.}
\newblock \bibinfo{journal}{\bibinfo{title}{Valley vortex states in sonic crystals}}.
\newblock {\emph{\JournalTitle{Physical Review Letters}}} \textbf{\bibinfo{volume}{116}}, \bibinfo{pages}{093901} (\bibinfo{year}{2016}).

\bibitem{lu2017observation}
\bibinfo{author}{Lu, J.} \emph{et~al.}
\newblock \bibinfo{journal}{\bibinfo{title}{Observation of topological valley transport of sound in sonic crystals}}.
\newblock {\emph{\JournalTitle{Nature Physics}}} \textbf{\bibinfo{volume}{13}}, \bibinfo{pages}{369--374} (\bibinfo{year}{2017}).

\bibitem{zhang2018topological}
\bibinfo{author}{Zhang, X.}, \bibinfo{author}{Xiao, M.}, \bibinfo{author}{Cheng, Y.}, \bibinfo{author}{Lu, M.-H.} \& \bibinfo{author}{Christensen, J.}
\newblock \bibinfo{journal}{\bibinfo{title}{Topological sound}}.
\newblock {\emph{\JournalTitle{Communications Physics}}} \textbf{\bibinfo{volume}{1}}, \bibinfo{pages}{97} (\bibinfo{year}{2018}).

\bibitem{laforge2019observation}
\bibinfo{author}{Laforge, N.} \emph{et~al.}
\newblock \bibinfo{journal}{\bibinfo{title}{Observation of topological gravity-capillary waves in a water wave crystal}}.
\newblock {\emph{\JournalTitle{New Journal of Physics}}} \textbf{\bibinfo{volume}{21}}, \bibinfo{pages}{083031} (\bibinfo{year}{2019}).

\bibitem{makwana2020experimental}
\bibinfo{author}{Makwana, M.~P.} \emph{et~al.}
\newblock \bibinfo{journal}{\bibinfo{title}{Experimental observations of topologically guided water waves within non-hexagonal structures}}.
\newblock {\emph{\JournalTitle{Applied Physics Letters}}} \textbf{\bibinfo{volume}{116}} (\bibinfo{year}{2020}).

\bibitem{laforge2021acoustic}
\bibinfo{author}{Laforge, N.} \emph{et~al.}
\newblock \bibinfo{journal}{\bibinfo{title}{Acoustic topological circuitry in square and rectangular phononic crystals}}.
\newblock {\emph{\JournalTitle{Physical Review Applied}}} \textbf{\bibinfo{volume}{15}}, \bibinfo{pages}{054056} (\bibinfo{year}{2021}).

\bibitem{WuNatComm2022}
\bibinfo{author}{Wu, X.} \emph{et~al.}
\newblock \bibinfo{journal}{\bibinfo{title}{{Topological phononics arising from fluidsolid interactions}}}.
\newblock {\emph{\JournalTitle{Nature Communications}}} \textbf{\bibinfo{volume}{13}}, \bibinfo{pages}{6120}, \doiprefix\url{https://doi.org/10.1038/s41563-025-02169-y} (\bibinfo{year}{2022}).

\bibitem{zhao:natmater2025}
\bibinfo{author}{Zhao, S.} \emph{et~al.}
\newblock \bibinfo{journal}{\bibinfo{title}{{Topological acoustofluidics}}}.
\newblock {\emph{\JournalTitle{Nature Materials}}} \textbf{\bibinfo{volume}{24}}, \bibinfo{pages}{707–715} (\bibinfo{year}{2025}).

\bibitem{eringen2003nonlocal}
\bibinfo{author}{Eringen, A.~C.} \& \bibinfo{author}{Wegner, J.}
\newblock \bibinfo{journal}{\bibinfo{title}{Nonlocal continuum field theories}}.
\newblock {\emph{\JournalTitle{Appl. Mech. Rev.}}} \textbf{\bibinfo{volume}{56}}, \bibinfo{pages}{B20--B22} (\bibinfo{year}{2003}).

\bibitem{landau2013electrodynamics}
\bibinfo{author}{Landau, L.~D.} \emph{et~al.}
\newblock \emph{\bibinfo{title}{Electrodynamics of continuous media}}, vol.~\bibinfo{volume}{8} (\bibinfo{publisher}{elsevier}, \bibinfo{year}{2013}).

\bibitem{krushynska2011normal}
\bibinfo{author}{Krushynska, A.~A.} \& \bibinfo{author}{Meleshko, V.~V.}
\newblock \bibinfo{journal}{\bibinfo{title}{Normal waves in elastic bars of rectangular cross section}}.
\newblock {\emph{\JournalTitle{The Journal of the Acoustical Society of America}}} \textbf{\bibinfo{volume}{129}}, \bibinfo{pages}{1324--1335} (\bibinfo{year}{2011}).

\bibitem{chen2024anomalous}
\bibinfo{author}{Chen, Y.} \emph{et~al.}
\newblock \bibinfo{journal}{\bibinfo{title}{Anomalous frozen evanescent phonons}}.
\newblock {\emph{\JournalTitle{Nature Communications}}} \textbf{\bibinfo{volume}{15}}, \bibinfo{pages}{8882} (\bibinfo{year}{2024}).

\bibitem{iglesias2025nonlocal}
\bibinfo{author}{Iglesias~Mart{\'\i}nez, J.~A.}, \bibinfo{author}{Chen, Y.}, \bibinfo{author}{Wang, K.} \& \bibinfo{author}{Wegener, M.}
\newblock \bibinfo{journal}{\bibinfo{title}{Nonlocal conduction in a metawire}}.
\newblock {\emph{\JournalTitle{Advanced Materials}}} \textbf{\bibinfo{volume}{37}}, \bibinfo{pages}{2415278} (\bibinfo{year}{2025}).

\bibitem{iglesias2021experimental}
\bibinfo{author}{Iglesias~Mart{\'\i}nez, J.~A.} \emph{et~al.}
\newblock \bibinfo{journal}{\bibinfo{title}{Experimental observation of roton-like dispersion relations in metamaterials}}.
\newblock {\emph{\JournalTitle{Science advances}}} \textbf{\bibinfo{volume}{7}}, \bibinfo{pages}{eabm2189} (\bibinfo{year}{2021}).

\bibitem{wang2022nonlocal}
\bibinfo{author}{Wang, K.}, \bibinfo{author}{Chen, Y.}, \bibinfo{author}{Kadic, M.}, \bibinfo{author}{Wang, C.} \& \bibinfo{author}{Wegener, M.}
\newblock \bibinfo{journal}{\bibinfo{title}{Nonlocal interaction engineering of 2{D} roton-like dispersion relations in acoustic and mechanical metamaterials}}.
\newblock {\emph{\JournalTitle{Communications Materials}}} \textbf{\bibinfo{volume}{3}}, \bibinfo{pages}{35} (\bibinfo{year}{2022}).

\bibitem{kazemi2023drawing}
\bibinfo{author}{Kazemi, A.} \emph{et~al.}
\newblock \bibinfo{journal}{\bibinfo{title}{Drawing dispersion curves: band structure customization via nonlocal phononic crystals}}.
\newblock {\emph{\JournalTitle{Physical Review Letters}}} \textbf{\bibinfo{volume}{131}}, \bibinfo{pages}{176101} (\bibinfo{year}{2023}).

\bibitem{fleury2014nonreciprocity}
\bibinfo{author}{Fleury, R.}, \bibinfo{author}{Sounas, D.~L.}, \bibinfo{author}{Sieck, C.~F.}, \bibinfo{author}{Haberman, M.~R.} \& \bibinfo{author}{Alù, A.}
\newblock \bibinfo{journal}{\bibinfo{title}{Sound isolation and giant linear nonreciprocity in a compact acoustic circulator}}.
\newblock {\emph{\JournalTitle{Science}}} \textbf{\bibinfo{volume}{343}}, \bibinfo{pages}{516--519} (\bibinfo{year}{2014}).

\bibitem{Goldsberry2022}
\bibinfo{author}{Goldsberry, B.~M.}, \bibinfo{author}{Wallen, S.~P.} \& \bibinfo{author}{Haberman, M.~R.}
\newblock \bibinfo{journal}{\bibinfo{title}{Nonreciprocity and mode conversion in a spatiotemporally modulated elastic wave circulator}}.
\newblock {\emph{\JournalTitle{Physical Review Applied}}} \textbf{\bibinfo{volume}{17}}, \bibinfo{pages}{034050}, \doiprefix\url{10.1103/PhysRevApplied.17.034050} (\bibinfo{year}{2022}).

\bibitem{galiffi2022photonics}
\bibinfo{author}{Galiffi, E.} \emph{et~al.}
\newblock \bibinfo{journal}{\bibinfo{title}{Photonics of time-varying media}}.
\newblock {\emph{\JournalTitle{Advanced Photonics}}} \textbf{\bibinfo{volume}{4}}, \bibinfo{pages}{014002--014002} (\bibinfo{year}{2022}).

\bibitem{palermo2020surface}
\bibinfo{author}{Palermo, A.}, \bibinfo{author}{Celli, P.}, \bibinfo{author}{Yousefzadeh, B.}, \bibinfo{author}{Daraio, C.} \& \bibinfo{author}{Marzani, A.}
\newblock \bibinfo{journal}{\bibinfo{title}{Surface wave non-reciprocity via time-modulated metamaterials}}.
\newblock {\emph{\JournalTitle{J. Mech. Phys. Solids}}} \textbf{\bibinfo{volume}{145}}, \bibinfo{pages}{104181} (\bibinfo{year}{2020}).

\bibitem{Torrent2018}
\bibinfo{author}{Torrent, D.}, \bibinfo{author}{Poncelet, O.} \& \bibinfo{author}{Batsale, J.-C.}
\newblock \bibinfo{journal}{\bibinfo{title}{Nonreciprocal thermal material by spatiotemporal modulation}}.
\newblock {\emph{\JournalTitle{Physical Review Letters}}} \textbf{\bibinfo{volume}{120}}, \bibinfo{pages}{125501}, \doiprefix\url{10.1103/physrevlett.120.125501} (\bibinfo{year}{2018}).

\bibitem{Kang2022}
\bibinfo{author}{Kang, J.} \& \bibinfo{author}{Haberman, M.~R.}
\newblock \bibinfo{journal}{\bibinfo{title}{Sound diffusion with spatiotemporally modulated acoustic metasurfaces}}.
\newblock {\emph{\JournalTitle{Applied Physics Letters}}} \textbf{\bibinfo{volume}{121}}, \doiprefix\url{10.1063/5.0097590} (\bibinfo{year}{2022}).

\bibitem{Goldsberry2025}
\bibinfo{author}{Goldsberry, B.~M.}, \bibinfo{author}{Norris, A.~N.}, \bibinfo{author}{Wallen, S.~P.} \& \bibinfo{author}{Haberman, M.~R.}
\newblock \bibinfo{journal}{\bibinfo{title}{Green’s function approach to model vibrations of beams with spatio-temporally modulated properties}}.
\newblock {\emph{\JournalTitle{Proceedings of the Royal Society A: Mathematical, Physical and Engineering Sciences}}} \textbf{\bibinfo{volume}{481}}, \doiprefix\url{10.1098/rspa.2024.0580} (\bibinfo{year}{2025}).

\bibitem{farhat2021spacetime}
\bibinfo{author}{Farhat, M.}, \bibinfo{author}{Guenneau, S.}, \bibinfo{author}{Chen, P.-Y.} \& \bibinfo{author}{Wu, Y.}
\newblock \bibinfo{journal}{\bibinfo{title}{Spacetime modulation in floating thin elastic plates}}.
\newblock {\emph{\JournalTitle{Phys. Rev. B}}} \textbf{\bibinfo{volume}{104}}, \bibinfo{pages}{014308} (\bibinfo{year}{2021}).

\bibitem{Koukouraki2024}
\bibinfo{author}{Koukouraki, M.}, \bibinfo{author}{Petitjeans, P.}, \bibinfo{author}{Maurel, A.} \& \bibinfo{author}{Pagneux, V.}
\newblock \bibinfo{journal}{\bibinfo{title}{Floquet scattering of shallow water waves by a vertically oscillating plate}}.
\newblock {\emph{\JournalTitle{Wave Motion}}} \doiprefix\url{10.2139/ssrn.5055878} (\bibinfo{year}{2024}).

\bibitem{Cassedy1963}
\bibinfo{author}{Cassedy, E.} \& \bibinfo{author}{Oliner, A.}
\newblock \bibinfo{journal}{\bibinfo{title}{Dispersion relations in time-space periodic media: {P}art i{\textemdash}{S}table interactions}}.
\newblock {\emph{\JournalTitle{Proceedings of the {IEEE}}}} \textbf{\bibinfo{volume}{51}}, \bibinfo{pages}{1342--1359}, \doiprefix\url{10.1109/proc.1963.2566} (\bibinfo{year}{1963}).

\bibitem{TessierBrothelande2023}
\bibinfo{author}{Tessier~Brothelande, S.} \emph{et~al.}
\newblock \bibinfo{journal}{\bibinfo{title}{Experimental evidence of nonreciprocal propagation in space-time modulated piezoelectric phononic crystals}}.
\newblock {\emph{\JournalTitle{Applied Physics Letters}}} \textbf{\bibinfo{volume}{123}}, \doiprefix\url{10.1063/5.0169265} (\bibinfo{year}{2023}).

\bibitem{Harwood2024}
\bibinfo{author}{Harwood, A.~C.} \emph{et~al.}
\newblock \bibinfo{journal}{\bibinfo{title}{Space-time optical diffraction from synthetic motion}}.
\newblock {\emph{\JournalTitle{Arxiv}}}  (\bibinfo{year}{2024}).
\newblock \eprint{2407.10809}.

\bibitem{Nassar2017}
\bibinfo{author}{Nassar, H.}, \bibinfo{author}{Xu, X.}, \bibinfo{author}{Norris, A.} \& \bibinfo{author}{Huang, G.}
\newblock \bibinfo{journal}{\bibinfo{title}{Modulated phononic crystals: Non-reciprocal wave propagation and {W}illis materials}}.
\newblock {\emph{\JournalTitle{Journal of the Mechanics and Physics of Solids}}} \textbf{\bibinfo{volume}{101}}, \bibinfo{pages}{10--29}, \doiprefix\url{10.1016/j.jmps.2017.01.010} (\bibinfo{year}{2017}).

\bibitem{Huidobro2021}
\bibinfo{author}{Huidobro, P.}, \bibinfo{author}{Silveirinha, M.}, \bibinfo{author}{Galiffi, E.} \& \bibinfo{author}{Pendry, J.}
\newblock \bibinfo{journal}{\bibinfo{title}{Homogenization theory of space-time metamaterials}}.
\newblock {\emph{\JournalTitle{Physical Review Applied}}} \textbf{\bibinfo{volume}{16}}, \bibinfo{pages}{014044}, \doiprefix\url{10.1103/physrevapplied.16.014044} (\bibinfo{year}{2021}).

\bibitem{Touboul2024}
\bibinfo{author}{Touboul, M.}, \bibinfo{author}{Lombard, B.}, \bibinfo{author}{Assier, R.~C.}, \bibinfo{author}{Guenneau, S.} \& \bibinfo{author}{Craster, R.~V.}
\newblock \bibinfo{journal}{\bibinfo{title}{High-order homogenization of the time-modulated wave equation: {N}on-reciprocity for a single varying parameter}}.
\newblock {\emph{\JournalTitle{Proceedings of the Royal Society A: Mathematical, Physical and Engineering Sciences}}} \textbf{\bibinfo{volume}{480}}, \doiprefix\url{10.1098/rspa.2023.0776} (\bibinfo{year}{2024}).

\bibitem{Zhu2020}
\bibinfo{author}{Zhu, X.} \emph{et~al.}
\newblock \bibinfo{journal}{\bibinfo{title}{Tunable unidirectional compact acoustic amplifier via space-time modulated membranes}}.
\newblock {\emph{\JournalTitle{Physical Review B}}} \textbf{\bibinfo{volume}{102}}, \bibinfo{pages}{024309}, \doiprefix\url{10.1103/physrevb.102.024309} (\bibinfo{year}{2020}).

\bibitem{Tirole2024}
\bibinfo{author}{Tirole, R.} \emph{et~al.}
\newblock \bibinfo{journal}{\bibinfo{title}{Second harmonic generation at a time-varying interface}}.
\newblock {\emph{\JournalTitle{Nature Communications}}} \textbf{\bibinfo{volume}{15}}, \doiprefix\url{10.1038/s41467-024-51588-z} (\bibinfo{year}{2024}).

\bibitem{Santini2024}
\bibinfo{author}{Santini, J.}, \bibinfo{author}{Pu, X.}, \bibinfo{author}{Palermo, A.}, \bibinfo{author}{Braghin, F.} \& \bibinfo{author}{Riva, E.}
\newblock \bibinfo{journal}{\bibinfo{title}{Controlling surface acoustic waves ({SAW}s) via temporally graded metasurfaces}}.
\newblock {\emph{\JournalTitle{Journal of Sound and Vibration}}} \textbf{\bibinfo{volume}{592}}, \bibinfo{pages}{118632}, \doiprefix\url{https://doi.org/10.1016/j.jsv.2024.118632} (\bibinfo{year}{2024}).

\bibitem{Ammari2024}
\bibinfo{author}{Ammari, H.}, \bibinfo{author}{Cao, J.}, \bibinfo{author}{Hiltunen, E.~O.} \& \bibinfo{author}{Rueff, L.}
\newblock \bibinfo{journal}{\bibinfo{title}{Scattering from time-modulated subwavelength resonators}}.
\newblock {\emph{\JournalTitle{Proceedings of the Royal Society A: Mathematical, Physical and Engineering Sciences}}} \textbf{\bibinfo{volume}{480}}, \doiprefix\url{10.1098/rspa.2024.0177} (\bibinfo{year}{2024}).

\bibitem{Taravati2022}
\bibinfo{author}{Taravati, S.} \& \bibinfo{author}{Eleftheriades, G.~V.}
\newblock \bibinfo{journal}{\bibinfo{title}{Microwave space-time-modulated metasurfaces}}.
\newblock {\emph{\JournalTitle{ACS Photonics}}} \textbf{\bibinfo{volume}{9}}, \bibinfo{pages}{305--318} (\bibinfo{year}{2022}).

\bibitem{Collis2014}
\bibinfo{author}{Collis, S.~S.}, \bibinfo{author}{Joslin, R.~D.}, \bibinfo{author}{Seifert, A.} \& \bibinfo{author}{Theofilis, V.}
\newblock \bibinfo{journal}{\bibinfo{title}{Issues in active flow control: {T}heory, control, simulation, and experiment}}.
\newblock {\emph{\JournalTitle{Progress in Aerospace Sciences}}} \textbf{\bibinfo{volume}{40}}, \bibinfo{pages}{237--289}, \doiprefix\url{https://doi.org/10.1016/j.paerosci.2004.06.001} (\bibinfo{year}{2004}).

\bibitem{Brunton2015}
\bibinfo{author}{Brunton, S.~L.} \& \bibinfo{author}{Noack, B.~R.}
\newblock \bibinfo{journal}{\bibinfo{title}{Closed-loop turbulence control: Progress and challenges}}.
\newblock {\emph{\JournalTitle{Applied Mechanics Reviews}}} \textbf{\bibinfo{volume}{67}}, \bibinfo{pages}{050801}, \doiprefix\url{10.1115/1.4031175} (\bibinfo{year}{2015}).
\newblock \eprint{https://asmedigitalcollection.asme.org/appliedmechanicsreviews/article-pdf/67/5/050801/6074501/amr\_067\_05\_050801.pdf}.

\bibitem{Online:Visser1993}
\bibinfo{author}{Visser, M.}
\newblock \bibinfo{title}{Acoustic propagation in fluids: an unexpected example of \uppercase{L}orentzian geometry} (\bibinfo{year}{1993}).
\newblock \eprint{gr-qc/9311028}.

\bibitem{Article:GiadaWiley}
\bibinfo{author}{Colombo, G.}, \bibinfo{author}{Palma, G.} \& \bibinfo{author}{Iemma, U.}
\newblock \bibinfo{journal}{\bibinfo{title}{Numerical aeroacoustic assessment of a metacontinuum device impinged by a laser-generated sound source}}.
\newblock {\emph{\JournalTitle{Journal of Engineering}}} \textbf{\bibinfo{volume}{2024}}, \bibinfo{pages}{1--11} (\bibinfo{year}{2024}).

\bibitem{Article:GiadaScientificReport}
\bibinfo{author}{Colombo, G.} \& \bibinfo{author}{Iemma, U.}
\newblock \bibinfo{journal}{\bibinfo{title}{Assessment of the performances and limitations of spacetime convective corrections for acoustic metacontinua design}}.
\newblock {\emph{\JournalTitle{Scientific Reports}}} \textbf{\bibinfo{volume}{15}}, \bibinfo{pages}{1--12} (\bibinfo{year}{2025}).

\end{thebibliography}
\end{document}